# Analysis of Prospective Flight Schemes to Venus Accompanied by an Asteroid Flyby

**Vladislav Zubko**[1]

[1] *Space Research Institute (IKI) of the Russian Academy of Sciences (RAS)*

This paper deals with the problem of constructing a flight scheme to Venus, in which a spacecraft flying to the planet after a gravity assist maneuver and transition to a resonant orbit in order to re-encounter with Venus, makes a passage of a minor celestial body. The 117 candidate asteroids from the NASA JPL catalogue, whose diameter exceeds 1 km, were selected. The flight trajectories which meet the criteria of impulse-free both flyby Venus and asteroid, and the subsequent landing on the surface of Venus were found within the interval of launch dates from 2029 to 2050. The trajectory of the spacecraft flight from the Earth to Venus including flyby of Venus and asteroids with a subsequent landing on the surface of Venus was analyzed.

## Introduction

The study of asteroids is an important part in the development of scientific knowledge of the composition and structure of the Solar System. Their study, which began with ground-based observations in the first half of the 19th century, has now moved into the stage of using space-based facilities (space telescopes, research probes). Among the most famous space probe missions, we can denote *Vega-1 and 2, Giotto, Deep Impact, NEAR, Hayabusa, Osiris - Rex* etc. In total, at the moment, about 35 small bodies of the Solar System have been studied by spacecrafts[1]. Among them, 7 were periodic comets, which is an extremely small number of studied objects, taking into account that modern catalogues include about 540 numbered comets and 620,000 asteroids [2]. Notice that currently a dozen more celestial bodies in the coming years (2022-2026) planned to be studied by spacecrafts, including a large main belt asteroid (16) Psyche, almost entirely composed of iron. The *Lucy* mission (it has been in flight since 2021) should make a great contribution to the study of the structure of Jupiter's Trojan asteroids. The *Stardust*, *Hayabusa*, *Hayabusa-2* missions developed space exploration by bringing to the Earth particles collected from celestial bodies: the

[1]From the list of missions to fly to asteroids given on the JPL website NASA https://ssd.jpl.nasa.gov/sb/targets.html ( Accessed 12/11/2022)
[2]According to the JPL website NASA https://ssd.jpl.nasa.gov/ (2022-12-22 17:07:03 UTC)

81P/Wild 2 comet (*Stardust*) and the 25143 Itokawa (*Hayabusa*), 162173 Ryugu (*Hayabusa 2*), 101955 Bennu (*Osiris - Rex*) asteroids.

Space missions involving direct exploration of one asteroid, i.e. approaching an asteroid with subsequent entry into its orbit or landing on its surface, have both an undeniable advantage, i.e. obtaining the maximum amount of data on the structure and composition of the asteroid, as well as some shortcomings, which include the insufficiency of the information obtained for a qualitative assessment of the structure and composition each of the minor celestial body of the Solar System. The latter result in the incompleteness of the global understanding of the nature, functioning and interaction of them at any of the stages of the evolution of the Solar System. It should be noted that an alternative approach is also possible when a spacecraft is launched, the purpose of which is to study a large number of asteroids by multiple approaches to them, for example, in the main belt region (as an example, the *Lucy* [1] and *Hannes* [2] missions concept).

A promising direction of the development of space missions seems to be including a stage of accompanying research as part of the main scientific program (i.g. studying asteroids). For example, in the *NEAR* [2,3] space mission to the (433) Eros asteroid the spacecraft also approached the (253) Matilda asteroid, which made it possible to determine the mass and obtain photos of surface of the latter. Another missions were *Galileo* [4] approached (243) Ida and (951) Gaspra, *Cassini-Huygens* approach (2685) Mazursky [5], *Ulysses* passed the gaseous tail of several comets, *New Horizons* [6] approached the (486958) Arrokot trans-Neptunian object in 2015.

The flyby of asteroids is being considered as a possible continuation for the missions currently operating; for example, for the spacecraft of the *Spectrum-Roentgen-Gamma* mission, studies have been carried out in [7,8], which show that after 2026 (the expected date of completion of the main part of the mission), a flyby of near-Earth asteroids is possible, such as (99942) Apophis, 1997 XF11 etc.

Let us highlight the works [9–11] in which the authors carried out the search for optimal flight trajectories to extreme trans-Neptunian objects, such as (90377) Sedna, 2012 VP113 using the Venus-Earth-Earth-Jupiter maneuvers and asteroids passages. Herewith, the authors found the asteroids which passage is possible at Earth-Earth and Earth-Jupiter sites, in particular, during the mission to Sedna in 2029 is possible to flyby the main belt asteroid (20) Massalia on Earth-Earth arc, and in 2034 an asteroid (16) Psyche on Earth-Jupiter arc. In [10] it was proposed to use a small spacecraft as a piggyback to the main mission, detaching it after the last gravity assist and sending two separate spacecraft simultaneously to two trans-Neptunian bodies.

In this paper, the scenario of a flight to Venus is studied, in which a spacecraft flying to the planet makes a passage of a minor celestial body after a gravity assist maneuver and transition to a resonant orbit[3] in order to re-encounter with Venus. Such scenario of a flight to Venus with gravity assist to deliver lander to a specific location on the planet's surface was proposed in [12] as a part of the development of a technique for constructing flight trajectories to Venus. The authors also showed the advantage of the obtained scenario in terms of attaining the required landing sites over the traditionally used approaches.

An interesting approach considering the flight to TNOs was proposed and studied in the work [13] to flyby simultaneously two TNOs (without separation unlike the case described earlier) within the framework of the one mission for example according to the research it is possible to flyby Huya and Quaoar if launching spacecraft in 2025.

The relevance of the study in this paper is due to the possibility of a significant expansion of the scientific research program of the mission to Venus using the described scenario provided landing in a given region of the planet's surface. Here, for some simplification, by landing in a given region we mean an impulse-free return to the planet after a flyby of the asteroid. The approach proposed in [13] can be extended to any mission to a planet with the subsequent transition to an orbit that is resonant to the planet's orbit.

---

[3] Under the spacecraft orbit resonant with the planet's orbit in the ratio $m$:$n$ (hereinafter, for brevity, we call such a spacecraft orbit resonant $m : n$ ), in this paper we understand the heliocentric spacecraft orbit, the ratio of the period of which to the period of the planet's orbit is a rational number $m / n$ .

As part of the study, 117 minor celestial bodies were selected that meet the main criterion, which is the size of studied the celestial body that should have been more than 1 km in average diameter[4]. Notice that to increase scientific value of possible mission and constrain our search the lower limit on diameter was taken 1 km. However, if restriction has been lowered to 0.5 km or 0.3 km the number of available asteroids will increase to 300 and 800 accordingly.

The flight trajectories which meet the criteria of an impulse-free flyby both Venus and an asteroid 35 out of 117 minor celestial bodies were found within the interval of launch dates from 2029 to 2050. Such interval is accepted as taking into account the fact that all the main missions for the exploration of Venus, namely, *Venera-D* [14], *DAVINCI +* [15], *VERITAS* [16], *EnVision* [17] planned to implementation in the period from 2029 to 2040s, another reason is that an interest to research Venus will not fade away for many more decades.

## 1. Determination of the spacecraft trajectory to the Venus with gravity assist and asteroid passages

Let's describe the main phases of constructing a spacecraft trajectory according to the Earth-Venus-Asteroid-Venus flight scheme. On heliocentric sections the trajectory is determined by the solution of the Lambert problem [20,21] and then patched together at the periapsis of the flyby hyperbolic trajectory of the spacecraft. Further algorithms and methods of calculating gravity assist maneuvers are based on the method of the patched conic approximation, as well as the method of calculating gravity assists outlined in [22-24].

Let us denote the arrival and departure heliocentric velocities of the spacecraft from the planet with the gravitational parameter $\mu_1$ as $\mathbf{V}^{-}, \mathbf{V}^{+}$, obtained from the Lambert problem; $\mathbf{u}$ - vector of the heliocentric velocity of the planet in the ecliptic frame.

---

[4] The estimation of object's diameter was taken from JPL web-site (url: https://ssd.jpl.nasa.gov/tools/sbdb_lookup.html#/ (Assesd 01.04.2023) considering asteroid absolute magnitude and albedo.

Knowing the heliocentric velocity of the spacecraft before and after the planet flyby and the planet's orbital velocity, the approach («–») and departure («+») planetocentric velocities of the spacecraft "at infinity" $\mathbf{V}^{-}{}_{r}, \mathbf{V}^{+}{}_{r}$, can be determined as follows:

$$\begin{cases} \mathbf{V}_r^{-} = \mathbf{V}^{-} - \mathbf{u} \\ \mathbf{V}_r^{+} = \mathbf{V}^{+} - \mathbf{u} \end{cases},$$

Let us determine an angle between $\mathbf{V}^{-}{}_{r}, \mathbf{V}^{+}{}_{r}$:

$$\alpha = \arccos \frac{V_r^{-} \cdot V_r^{+}}{V_r^{-} V_r^{+}}, \tag{1}$$

where $V_r^{-} = |\mathbf{V}_r^{-}|$, $V_r^{+} = |\mathbf{V}_r^{+}|$. On the other hand, assuming that the incoming and outgoing asymptotes are patched at the pericenter of the spacecraft trajectory, the expression for the turn angle can be written as:

$$\alpha = \arcsin \frac{1}{1 + r_\pi \left(V_r^{-}\right)^2 / \mu_1} + \arcsin \frac{1}{1 + r_\pi \left(V_r^{+}\right)^2 / \mu_1}, \tag{2}$$

where $r_\pi$ is the magnitude of the pericenter radius of the spacecraft's flyby trajectory.

The resulting nonlinear equation is solved with respect to the variable $r_\pi$, with the required rotation angle α known. From (2) the pericenter radius at which the tangents to both semihyperbolas coincide is determined.

The condition under which the gravity assist maneuver is possible only due to the gravitational field of the planet (i.e. passive or free-impulse gravity maneuver) is written as follows:

$$V_r^{-} = V_r^{+},\ \alpha \le \alpha^{*}, \tag{3}$$

where $\alpha^{*} = \arcsin \frac{1}{1 + r_{\pi,\min} \left(V_r\right)^2 / \mu_1}$ is the maximum natural turn angle, $r_{\pi,\min}$ is the minimal planetocentric distance at which flyby can be occurred, for example, in our case, for Venus such distance was calculated as the sum of the average Venus radius (6051 km) and the average height of Venusian atmosphere (500 km), thus $r_{\pi,\min} = 6551\ km$.

If the first of the conditions (3) is not satisfied, then an additional impulse is required. Assuming that the impulse is applied at the pericenter of the spacecraft's flyby trajectory, we can determine its value as follows:

$$\Delta V_{\pi} = \sqrt{\frac{2\mu_1}{r_{\pi}} + \left(V_r^-\right)^2} - \sqrt{\frac{2\mu_1}{r_{\pi}} + \left(V_r^+\right)^2}\ . \tag{4}$$

Note that such impulse is not exactly optimal: as it was shown in papers [22,25], the optimal impulse should be applied not in the pericenter of the flyby trajectory. These papers give formulas for calculating the distance at which the transition between semihyperbolas should be made and the impulse required for such a transition. In our study, since the impulse calculated by eq. (4) will be close to the optimum in [22,25], for simplification of calculations we use exactly such approach. Fig. 1 shows the calculation scheme of the active gravity assist maneuver (i.e., with a nonzero impulse as in eq. (4)) when passing the planet.

If the second of the conditions (3) is not fulfilled, then another impulse is required. That impulse required to turn the asymptote entering or leaving the sphere of influence of the planet so that the condition (2) is fulfilled. The value of such impulse is defined as follows:

$$\Delta V_t = 2V_r^- \sin\frac{\Delta\alpha}{2}, \tag{5}$$

where $\Delta\alpha = \left|\alpha - \alpha^*\right|$ is an angle by which the asymptote was turned.

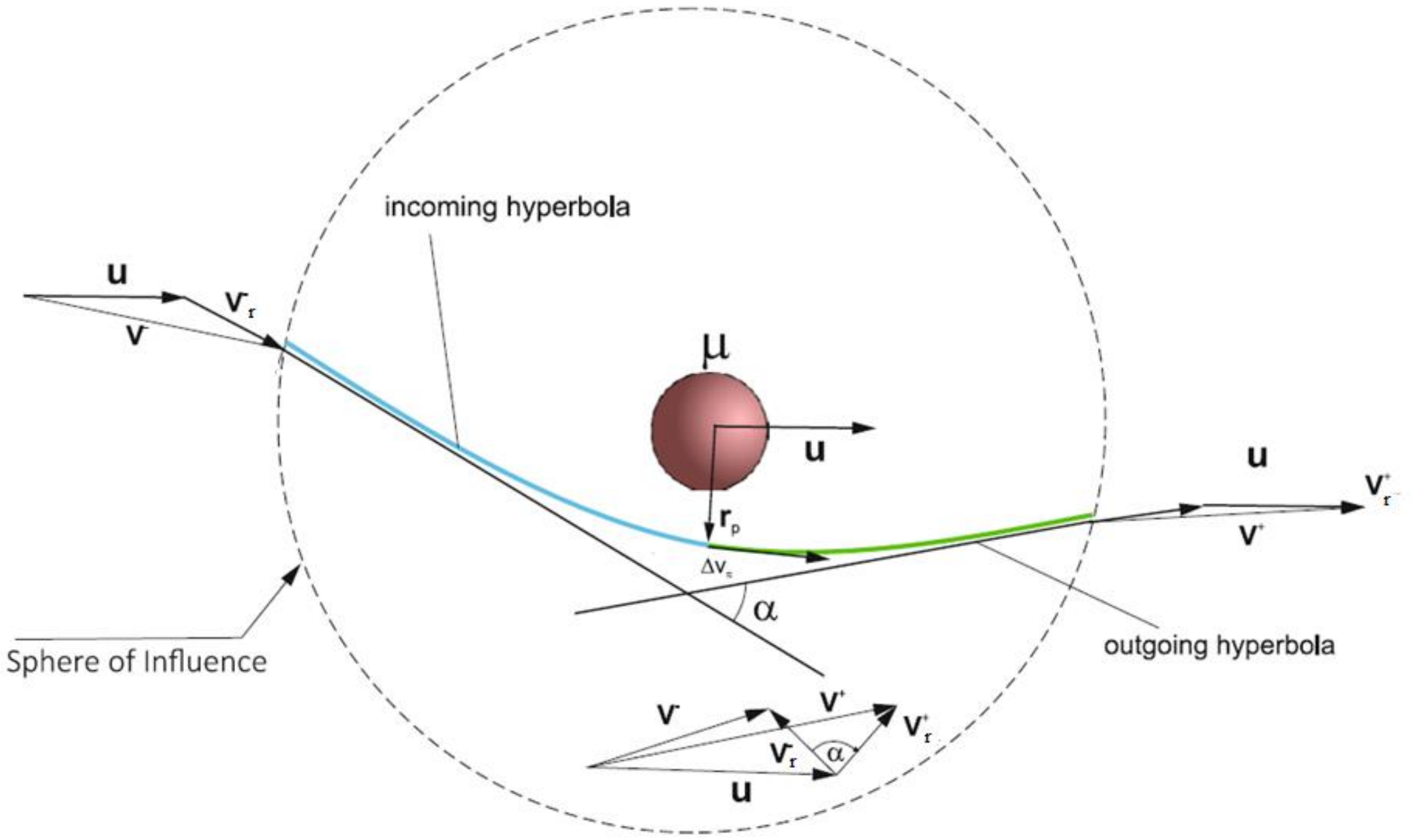

Fig. 1. Scheme of constructing the trajectory of the spacecraft during the passage of the planet stage in the case of an active gravity assist maneuver

## 2. Technique for calculation of the trajectory of the spacecraft from the Earth to Venus including flyby of Venus and asteroids with a subsequent return to it

The technique used in the study of flight trajectories can be outlined as follows:

1. The celestial bodies Earth, Venus, and asteroids, which are to be orbited during the motion of the spacecraft along the resonant orbit of a given resonance 1:1, are specified. The coordinates of the given celestial bodies are calculated for the time moments corresponding to the date of launch the spacecraft, *p* passages of the celestial bodies and landing on Venus.
2. Sections of the spacecraft trajectory are calculated: Earth-Venus-1$^{st}$ asteroid-2$^{nd}$ asteroid-...-$p^{th}$ asteroid-Venus, using the solution of the Lambert problem[5] for each section (Fig. 2). The result of this stage of the technique are vectors of heliocentric velocities of the spacecraft at each of the moments of encounter with the given celestial body.

[5] In this work we used Izzo method described in details in [31] and verified results using Sukhanov method [32].

3. Based on the data obtained in the previous step, the impulses are calculated $\Delta V_{ai} = \Delta V_{\pi i} + \Delta V_{ti},\ i = 0,\dots,p$ (eq. (4) – (5)), assuming that all gravity assist maneuvers are active (i.e. require an additional impulse during flyby). Impulse with the index «0» corresponds to the first gravity assist maneuver, i.e. near Venus. Also, we assume that during an asteroid flyby the trajectory of the spacecraft does not experience gravitational influence from the asteroid. Thus, the above formula is simplified to calculate the modulus of the difference of asymptotic velocities before and after the asteroid flyby:

$$\Delta V_{ai} = |\mathbf{V}_{rai}^{-} - \mathbf{V}_{rai}^{+}|,\ i = 1,\dots,p,$$

where $\mathbf{V}_{ra}^{-}, \mathbf{V}_{ra}^{+}$ are asymptotic velocities before and after the flyby of the asteroid.

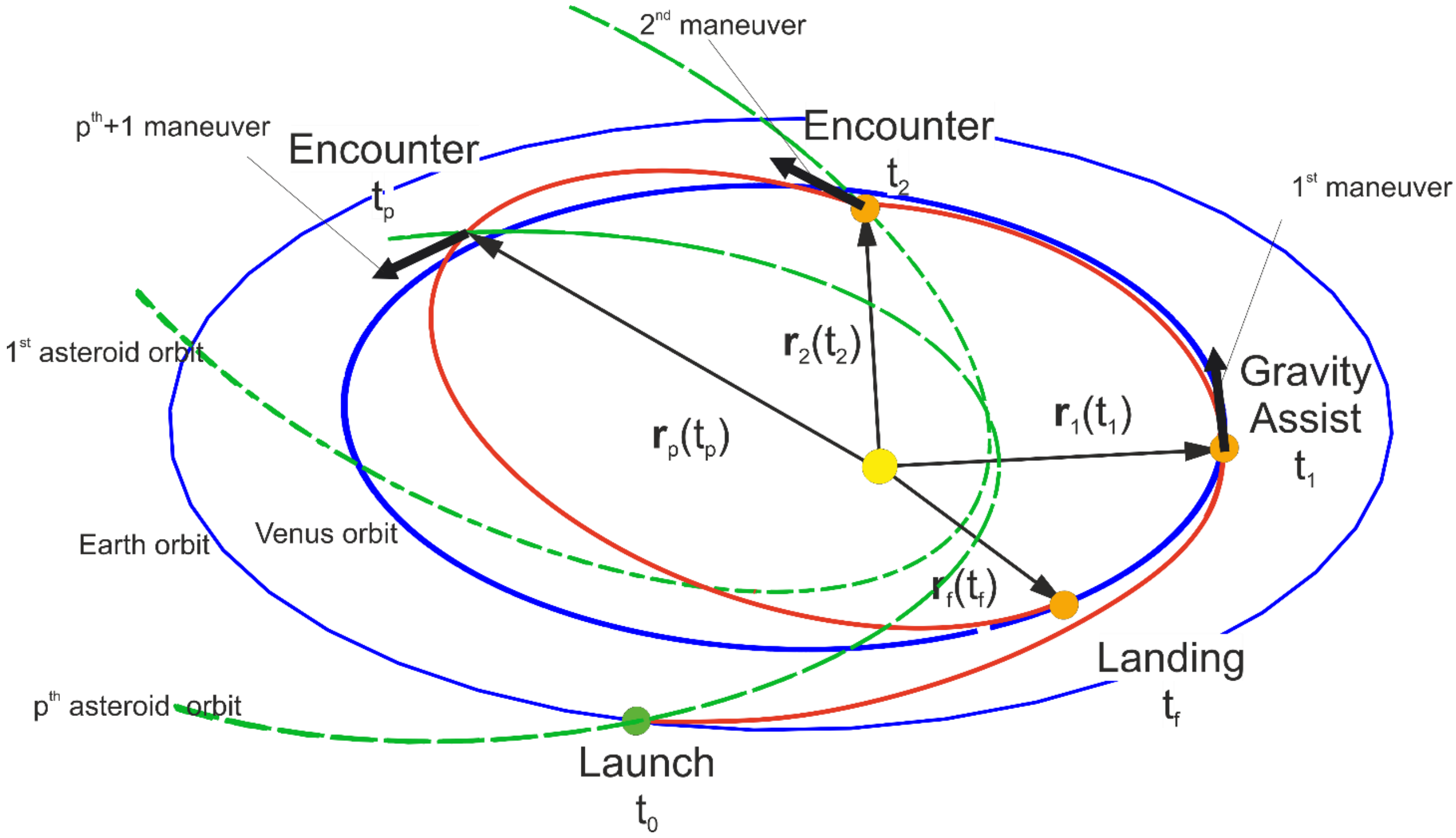

Fig. 2. Stage of mssion scenario acording to used technique. $\mathbf{r}_1, \mathbf{r}_2, \mathbf{r}_p, \mathbf{r}_f$ are the position of spacecraft at Venus first gravity assist, asteroids gravity assist and final position at landing

4. The total impulse $\Delta V_{\Sigma}(\mathbf{T})$, where $\mathbf{T} = \{t_0, t_1, \dots, t_i, \dots, t_p, t_f\}$, can be calculated, as sum of $p$ impulses, the asteroid fly-bys applied by the spacecraft $\Delta V_{ai}(t_i, t_{i+1}),\ i = 1,\dots,p$, as well as launch characteristic velocity $\Delta V_0(t_0,t_1)$ [6], i.e. $\Delta V_\Sigma(\mathbf{T}) = \sum_{i=1}^{p} \Delta V_{ai}(t_i,t_{i+1}) + \Delta V_0(t_0,t_1)$.

5. The total impulse is minimized by the criterion $J = \Delta V_\Sigma(\mathbf{T}) \underset{\mathbf{T}}{\rightarrow} \min$ using any available method of finding the minimum of the objective function *J*. Notice that in this research *J* was minimized using the two-step optimization procedure with a differential evolution on the first step and the Broyden-Fletcher-Goldfarb-Shanno method on the second step. The implementation of the mentioned methods was made according to [18,19].
6. Then the trajectories are filtered according to the following criterion:

$$\begin{cases} \tau_1 = t_2 - t_1,\ t_2 > t_1 \\ \tau_2 = t_3 - t_2,\ t_3 > t_2 \\ \ldots \\ \tau_p = t_{p+1} - t_p,\ t_{p+1} > t_p \end{cases}$$

$$\left\| \sum_{i=1}^{p} \tau_i - T_{pl} \right\| \le \varepsilon ,$$

where $T_{pl}$ — planet orbital period, $\varepsilon$ — the required accuracy.
7. If the duration of the spacecraft flight along the planet-asteroid-planet section does not satisfy the above inequality, it is considered that the passage of the asteroid during the simultaneous flight on the turn of the resonant orbit is impossible and the above mentioned steps 1-6 should be repeteaed for a new launch date.

## 3. Calculations

The candidate asteroids for current research were selected from the JPL catalog in the Horizons system. The criteria for selecting candidate asteroids from the JPL catalog are:

- perihelion < 0.8 au;
- object diameter > 1 km.

[6] Launch characteristic velocity was calculated considering that start of spacecrafts motion occurs from Earth from the parking orbit of 200 km altitude and 51.6 deg inclination. The equation for required impulse calculation may be found in [33]

The first criterion is based on the fact that the orbit of the asteroid should be passing close to the orbit of Venus. Such a requirement arises because of the limited region of space where the spacecraft can be sent after being placed on the resonant 1:1 orbit. Estimation of the maximum distance to which the spacecraft can leave after a resonant 1:1 orbit after a gravity assist of Venus can be made in the simplest model using the Tisserand invariant which can be written through the orbital parameters and the hyperbolic excess velocity as

$$T = \frac{a_{pl}}{a} + 2\sqrt{\frac{a}{a_{pl}}(1-e^2)}\cos i\,,$$

$$T = 3 - \left(\frac{V_r}{u}\right)^2,$$

where $a, a_{pl}$ - the semi-major axis of the spacecraft orbit after the gravity assist maneuver and the Venus orbit; $e, i$ - eccentricity and inclination of the heliocentric orbit of the spacecraft in the plane of the orbit of Venus.

Then the maximal distance in the aphelion ($r_{\alpha,\max}$) is achieved if the orbit of the spacecraft lies in the plane of the orbit of Venus, i.e. $i = 0$ and $e = e_{\max}$ ($e_{\max}$ is the maximal eccentricity of heliocentric orbit of the spacecraft), also take into account that the spacecraft after the gravity assist maneuver transfers to the resonant 1:1 orbit, i.e. $a = a_{pl}$:

$$e_{\max} = \sqrt{1 - \frac{1}{4}\left(2 - \left(\frac{V_r}{u}\right)^2\right)^2}\,,$$

$$r_{\alpha,\max} = a(1 + e_{\max})\,.$$

Hence, we see that for the asymptotic velocities within the range from 3 to 15 km/s the maximal aphelion in the resonant 1:1 orbit which the spacecraft reaches after the gravity assist maneuver near Venus varies from 0.78 to 1 au. Thus, choosing the first criterion as the upper limit of perihelion value of the candidate asteroid orbit allows to remove those asteroids which basically cannot be reached by spacecraft flying on the resonant 1:1 orbit.

The second criterion allows us to identify only relatively large objects, which may be of greater interest for scientific research. A total of 117 minor celestial bodies was found, of which 4 are comets that meet the above criteria.

First, we investigated the case of an impulse-free asteroid flyby spacecraft while flying by the Earth-Venus heliocentric arc. In this case, the trajectory of the spacecraft in this arc can be found by solutions of the Lambert problem at the Earth-Asteroid and the asteroid-Venus arcs. However, when searching among 117 asteroids in the interval of launch dates from 2029 to 2050, we were unable to find an asteroid which passage could be carried out, firstly, without an impulse, and secondly, within the duration of the flight corresponding to the limiting time of the spacecraft flight along the trajectories of the second semi turn[7].

Further, according to the technique given in section 2, simulations of the trajectories of the spacecraft flight following flyby Venus and the asteroid, and the subsequent landing on the surface of Venus were carried out. The main constraints to the trajectories of spacecraft flight were the following parameters, namely the launch characteristic velocity and impulse-free of the passage of both Venus and the asteroid:

$$\begin{cases} \Delta V_0 \leq 4.5\ km/s \\ \Delta V_{ai} = 0, \quad i = 0, ..., p \end{cases} \qquad (6)$$

Note that the choice of the first of the constraints (6) is due to the fact that usually the characteristic velocity value of the spacecraft for Venusian missions ranges from ~3.6 to ~4.2 km/s [14,20]. In our case, we increased this limit up to 4.5 km/s in order to expand the number of available scenarios. Table A.1 also includes the scenario of a flight to Venus circling the 2P/Encke comet, due to the high scientific value of studying the comet at a close range, despite the fact that the restrictions on the launch characteristic velocity are violated (5.25 km/s vs. 4.5 km/s in constraint (6)). This case is considered below in detail.

[7] By the first and second semi-turn usually refers to the flight of spacecraft from Earth to Venus on the trajectory, providing a rendezvous with Venus at an angle distance of less than 180 degrees (the first semi-turn) or more than 180 degrees (the second semi-turn).

The analysis of the possibility of an asteroid passage was carried out on the interval of possible years of launch missions to Venus 2029-2050[8]. Asteroid ephemeris was obtained using NASA's JPL HORIZONS catalog of small celestial bodies[9]. In the appendix, in Table A.1 some characteristics of the flight to Venus with the gravity assist maneuver and the subsequent passage of the asteroid are given. Fig. 2*a* and 2*b* show the main characteristics of certain trajectories, which include an impulse-free encounter with the asteroid. Fig. 3*a* shows launch characteristic velocity ($\Delta V_0$) for each of the asteroids from the JPL catalog, the passage of which is possible under the constraints (6), Fig. 2*b* shows *Vr* vs. the id of asteroid in JPL catalogue.

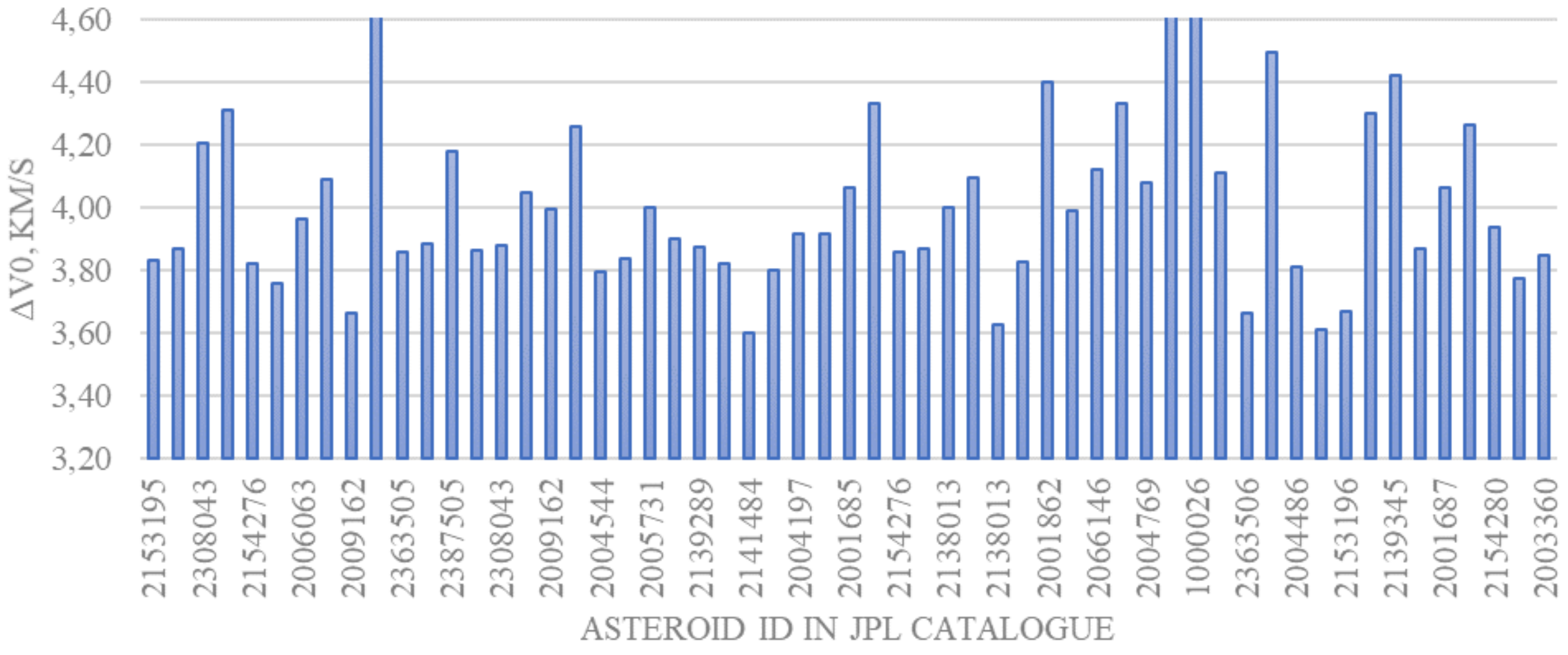

(a)

[8] It is worth clarifying that exploration missions to Venus are planned for the late 2020s to mid-2030s. However, we are referring here to possible post-2030 missions.

[9] JPL Small-body database https://ssd.jpl.nasa.gov/tools/sbdb_lookup.html#/ (accessed 10.12.2022)

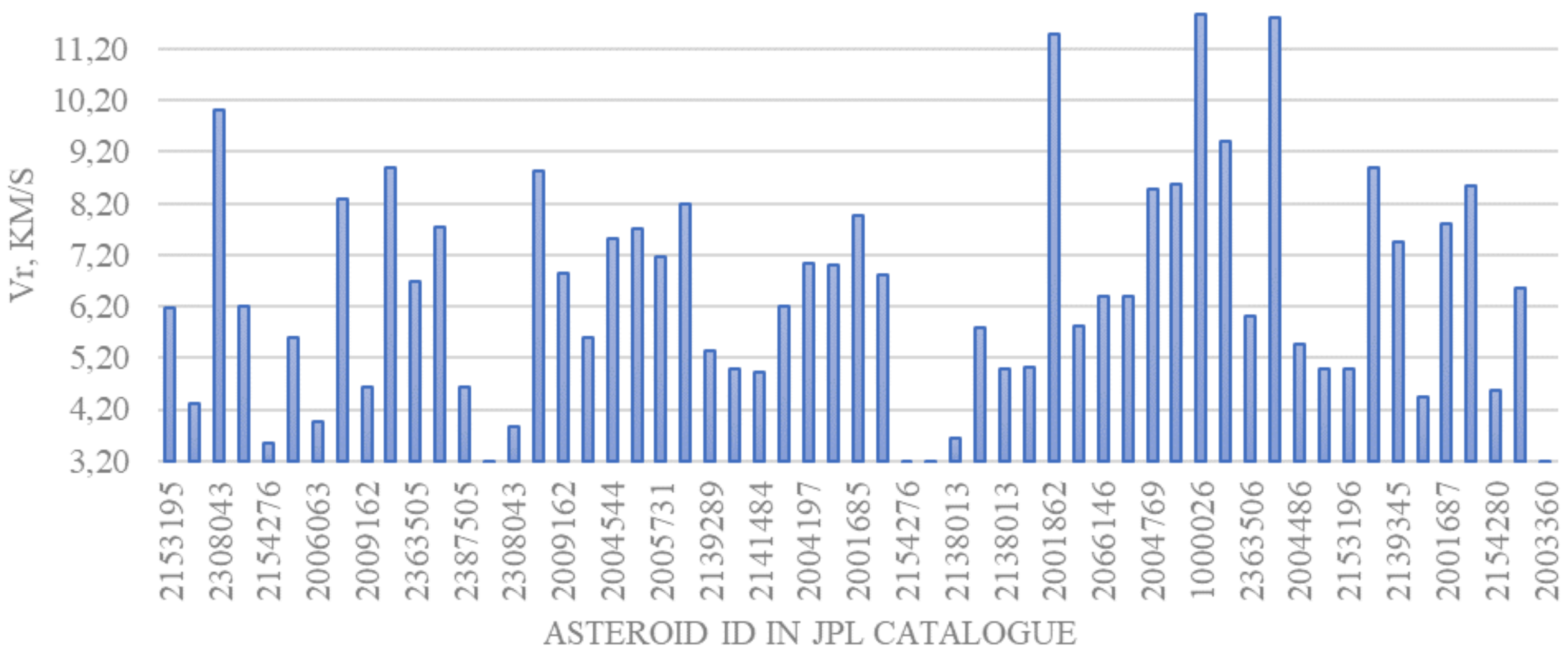

(b)

Fig. 3 (a) Distribution of $\Delta V_0$ vs. asteroid id in the JPL catalog, flight to which is possible under the constraints (6) (see Table A.1); (b) Distribution of $V_r$ vs. asteroid id in the JPL catalog, flight to which is possible within the constraints (6) (see Table A.1)

Fig. 3 does not show the values of $\Delta V_0$ and $V_r$ for the scenario with a flyby of the 2P/Encke comet, in spite of the fact that these results are shown in Table A.1. This is because the optimal $\Delta V_0$ obtained is significantly higher than the first part of the constraint (6) introduced earlier.

Note that in the process of searching for optimal trajectories the following circumstance was detected: the optimal in $\Delta V_0$ term trajectories satisfying the condition (6) were spacecraft trajectories on which the duration of the flight on the Venus-asteroid-Venus section was equal to the period of the planet. This means that the optimal trajectory turned out to be that which would contain a section of the spacecraft flight along the resonant *1:n* orbit in relation to the orbit of Venus.

Results presented in Table A.1 shows that the conclusion about the optimality is violated only in two cases, namely the flight to the 2P/Encke comet in 2042, whereas the flight along the 1:2 orbit in the Venus-asteroid-Venus arc turns out to be optimal, the other case is the meeting with the 6063 Jason (1984 KB) asteroid at launch in 2034, then the flight along the Venus-asteroid-Venus arc of 309 days is optimal. However, the latter may be explained that the spacecraft flight after the gravity assist maneuver takes place in the Venus orbital plane.

Also, it should be pointed out that the above conclusion about optimality can be valid if the spacecraft meets the asteroid after Venus flyby occurs at an angular distance of up to $2\pi$ (on heliocentric arc).

Although the condition for impulse-free solution for flyby both Venus and asteroid was noticed and established during the simulation of flight paths to 117 asteroids in the study of the launch date interval from 2029 to 2050, but strong evidence or explanation of this circumstance has not been succeeded yet, and this is the subject of further research.

## 4. Analysis of the results

We found 53 possible scenarios of spacecraft flight to Venus with an asteroid flyby, satisfying constraints (6) and 3 scenarios, with flyby of the 2P/Encke comet (see Table A.1). Fig. 4 shows the distribution of the number of possible scenarios vs. launch year.

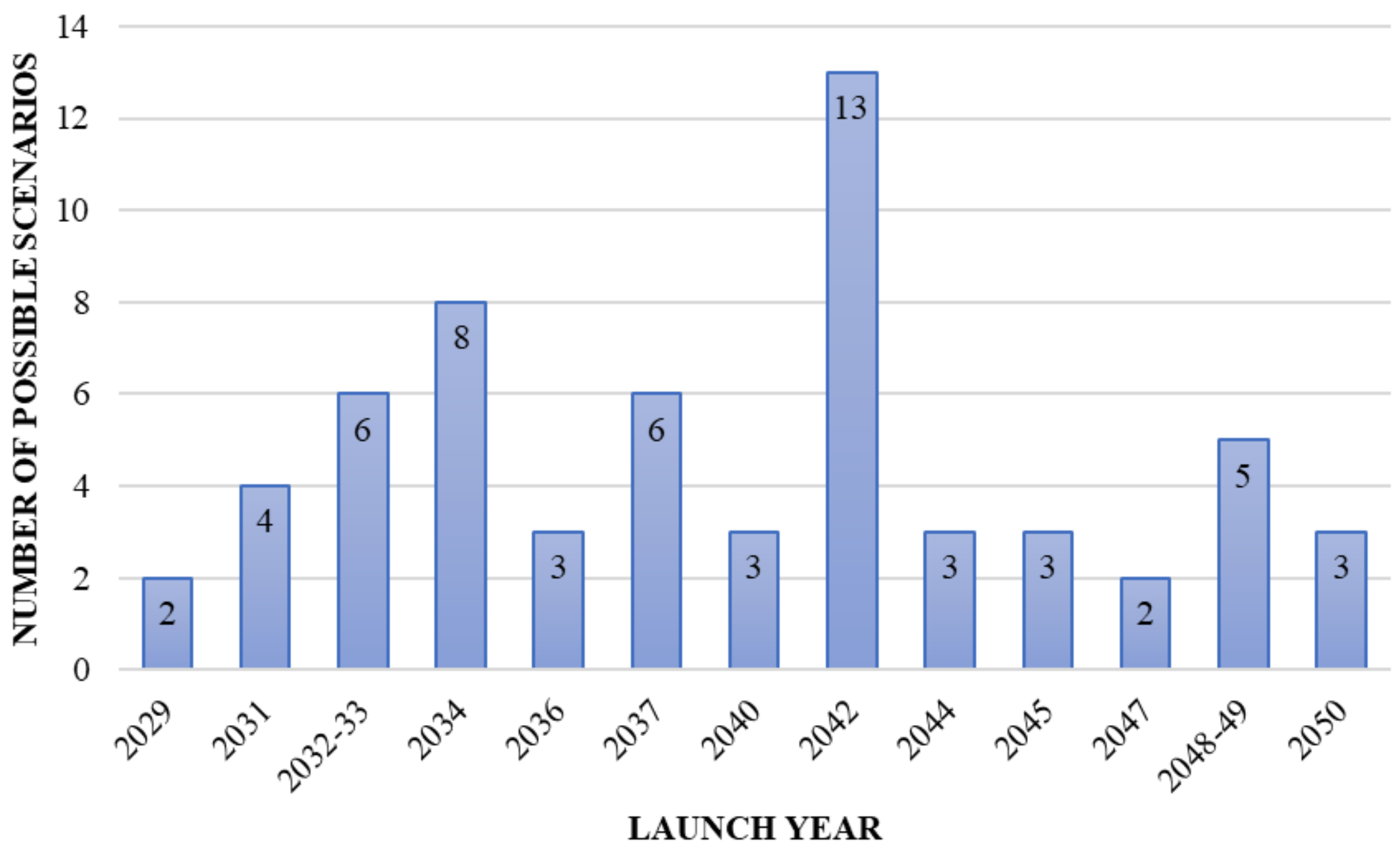

Fig. 4 Distribution of the number of possible scenarios of asteroid flyby on the Venus-Venus trajectory for various years of mission launch

From this figure one can see that the number of possible scenarios varies from 2 in 2029, 2040, 2044-2047 to 12 in 2042, so, in general, in all considered years of launch the spacecraft to Venus it is possible to find at least one scenario including an asteroid impulse-free flyby.

Let us analyze the sizes of the objects, the passage of which is possible in the framework of the proposed scenario of the mission of spacecraft to Venus (see Table A.1). For this purpose, it is necessary to demonstrate the asteroid's diameter[10] vs. its number in the JPL catalog (see Fig. 5).

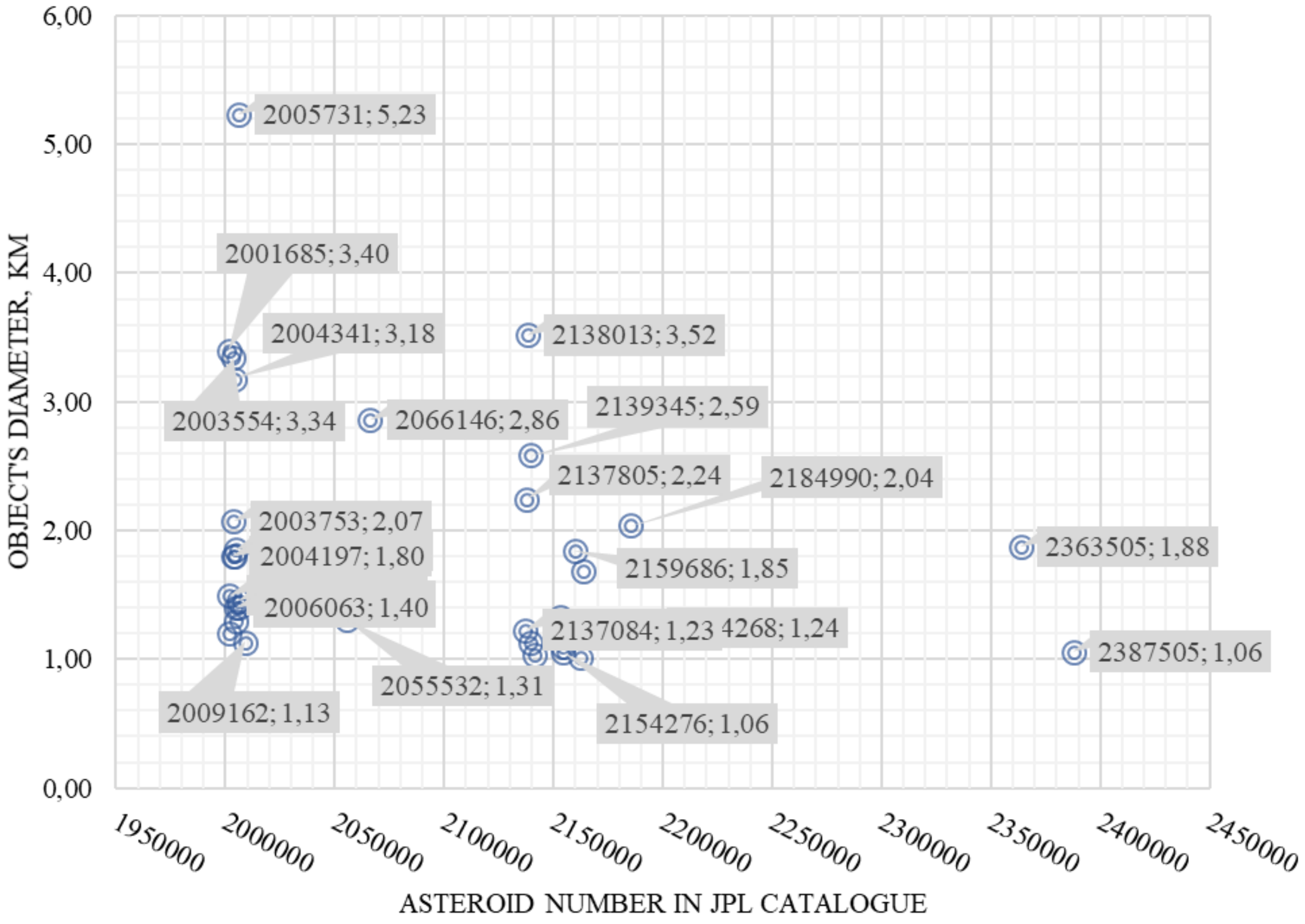

Fig. 5 Object's diameter vs. the number of reachable asteroids on the Venus-Venus trajectory arc (the diagram in the format (xxxxxxxx; x,xx) shows the asteroid id in the JPL catalog and its diameter in km).

The results of search for trajectories including flyby of Venus and asteroid in launch dates period from 2029 to 2050 show that most of the found objects are relatively small and have sizes that ranges from 1 km (corresponds to the lower border) to 5.23 km. The largest one that has been found is the 5731 Zeus asteroid

[10] The estimation of object's diameter was taken from JPL web-site (url: https://ssd.jpl.nasa.gov/tools/sbdb_lookup.html (Assesd 01.04.2023) considering asteroid absolute magnitude and albedo.

(5.23 km in diameter), this object may be reached at launch in 03.06.2036 and a close encounter occurs at 27.10.2036 at a relative velocity of ~14 km/s.

It is worth noting that for a number of asteroids the taxonomic type determined from the Earth observations is shown in Table 1. Analysis of these data shows that even at a very close distances to the Sun (0.8-1.12 au) one can make passage of quite interesting objects, such as the M-class asteroid 3554 Amun, Q-class asteroid 1862 Apollo, 3753 Cruithne. Asteroid 3554 Amun is notable because it has an own satellite, while 3753 Cruithne is in orbital 1:1 resonance with the Earth. Also, a 4769 Castalia asteroid which periodically approaches Earth at distances up to ~0.02 au (the next such approach is expected in 2047) is of interest because it belongs to the contact-double S-class. It is also worth paying attention to the 2004 TG10 asteroid which supposedly could be a fragment of the 2P/Encke comet [21] and is also approaching the Earth up to 0.02 au.

Table 1. Physical properties (diameter, magnitude and taxonomic type) of some asteroids from Table A.1, according to [22] and the JPL catalog of small celestial bodies

| Number and name of the object | Launch year | H* | albedo | d*, km | Taxonomy class |
|---|---|---|---|---|---|
| 4341 Poseidon (1987 KF) | 2031 | 16.08 | 0.05 | 3.17 | O |
| 4197 Morpheus (1982 TA) | 2037 | 15.03 | 0.37 | 1.80 | Sq |
| 1862 Apollo (1932 HA) | 2042 | 16.11 | 0.25 | 1.50 | Q |
| 55532 (2001 WG2) | 2042 | 16.15 | 0.37 | 1.30 | Sk |
| 3753 Cruithne (1986 TO) | 2047 | 15.5 | 0.36 | 2.07 | Q |
| 5731 Zeus (1988 VP4) | 2036 | 15.43 | 0.03 | 5.23 | X[22] |
| 3554 Amun (1986 EB) | 2040, 2042, 2044 | 15.91 | 0.07 | 3.34 | X; M;D |
| 3360 Syrinx (1981 VA) | 2050 | 15.9-17.4 | 0.07-0.17 | 1.80 | D |
| 1685 Toro (1948 OA) | 2032, 2040, 2048-49 | 14.31 | 0.26 | 3.75 | S [22] |
| 4769 Castalia (1989 PB) | 2042 | 16.9 | 0.3 | 1.40 | S [22] |
| 2004 TG10 | 2031, 2034 | 19.4 | 0.018 | 1.36 | C, S[21] |

*H, d - absolute magnitude and diameter of the asteroid

## 5. **Flight scenario to Venus with a flyby of the 2P/Encke comet**

Due to the exclusivity of both the 2P/Encke comet [23–25] and the fact that 2P/Encke is the only comet among the 35 objects to which it was possible to construct trajectories, so let us consider this flight scenario in detail.

The flight to the 2P/Encke comet after the first flyby of Venus, according to our calculations, can be made at launch in 2033 or 2042. We consider 2033 since it is the most possible date of implementation of the mission to Venus and the flight to the comet will be accompanied by a lower cost $\Delta V_0$ than launched in 2042. The trajectory parameters of the flight to Venus with the passage of the 2P/Encke comet are given in Table A.1. In Table 2 the parameters of the resonant orbit during the motion on which spacecraft encounters the comet are provided.

Table 2. Orbital parameters of the resonant orbit of the spacecraft approaching the 2P/Encke comet after the gravity assist of Venus at 02.06.2033

| *a, au* | *e* | *i*, deg* | *Ω*, deg* | *ω*, deg* | *M**, deg* |
|---|---|---|---|---|---|
| 0.723 | 0.036 | 12.2 | 308.870 | 246.614 | 98.538 |

* in the heliocentric ecliptic frame, ** mean anomaly is given at the moment of the spacecraft encounter with the comet

Fig. 6 shows the trajectory of the spacecraft flight to Venus, with a subsequent impulse-free encounter with the 2P/Encke comet and return to Venus. It can be seen that the spacecraft moving along the resonant orbit passes the comet out of the ecliptic plane, which is primarily explained by the fact that the descending node of the orbit of the comet is located inside the Mercury orbit, and the ascending one is near the Jupiter orbit, thus make it impossible for the spacecraft to encounter 2P/Encke in the ecliptic plane and then return to Venus performing the flight on the resonant 1:1 orbit.

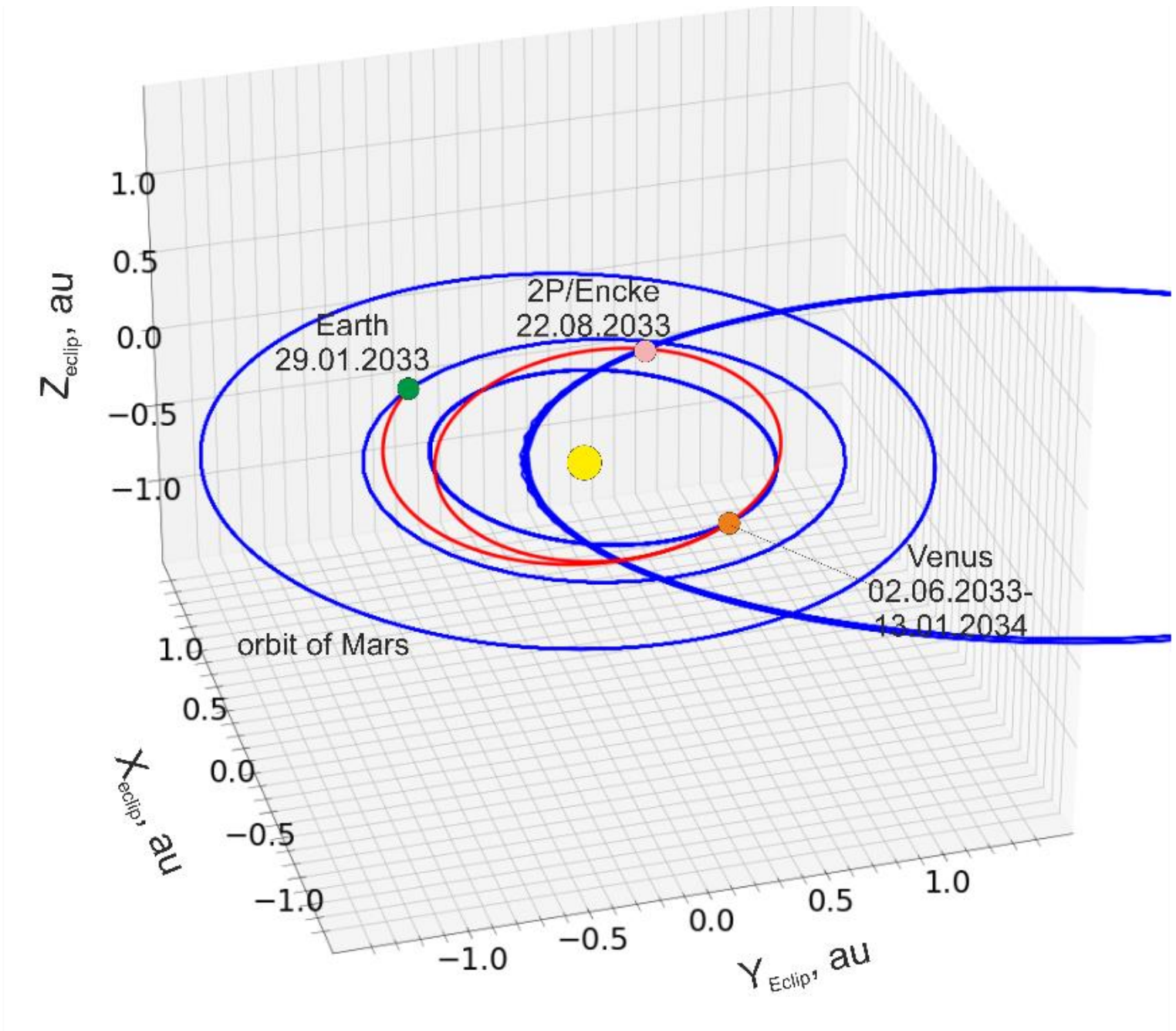

Fig. 6. Spacecraft's trajectory to Venus with the gravity assist maneuver near the planet and flyby of the 2P/Encke comet in a resonant 1:1 orbit at mission launch in 29.01.2033. Dates are written in the format dd.mm.yyyy.

Note that the required $\Delta V_0$ cost for a flight to Venus with its passage around, transition to the resonant 1:1 orbit, and an impulse-free encounter with the 2P/Encke comet is 5.13 km/s at launch in 29.01.2033. This value of characteristic velocity exceeds the typical $\Delta V_0$ required for Venusian missions, which is in the range of ~3.6-4.2 km/s [14,20], and less surpasses the optimal $\Delta V_0$ = 4.24 km/s for the direct flight from the Earth to the 2P/Encke comet in the nearest launch date 04.10.2029 with a flight duration of ~287 days. The explanation of the high value of launch characteristic velocity for the Earth-Venus-Encke-Venus flight may be the necessity of making a transition to an orbit with a great inclination to ecliptic, as the resonant orbit of the spacecraft after gravity assist maneuver near Venus is inclined at 12.2 degrees to ecliptic (Table 2) or at 14.47 degrees to the plane of the Venus orbit. For this spacecraft it is required to approach Venus with high asymptotic velocity, which follows from a simple relation:

$$\Delta i = \arcsin \frac{V_r \sin \Phi}{u},$$

where $\Phi$ is the angle between the vectors $\mathbf{u}, \mathbf{V}_r$, for a resonance of 1:1 $\sin \Phi \approx 0.997564 \approx 1$. Taking this into account, the possible limits of changing the inclination by means of a gravity assist maneuver are no more than $\Delta i = 14.5$ deg (for the asymptotic velocity of 8.88 km/s (Table A.1)). Because of the latter, and considering the requirement introduced by impulse-free passage both Venus and an asteroid, the spacecraft performs a flight, initially being outside the ecliptic plane (10.7 degrees of inclination of the spacecraft orbit on the Earth-Venus section). This naturally leads to an increase of $\Delta V_0$, but at the same time provides the fulfillment of the second part of the constraints (6).

Note that the reduction of $\Delta V_0$ within this scenario occurs if we increase the flight time along the Venus-Venus section. In this case the optimal $\Delta V_0$ solution for the resonant 1:2 orbit (Fig. 6) may be obtained, in this case $\Delta V_0$ is 4.65 km/s at launch in 30.05.2042 (142 days flight between the Earth and Venus), the encounter with the comet occurs on 17.10.2043 at the relative velocity 36.7 km/s. However, even in this scenario $\Delta V_0$ is still high and exceeds the established first of the constraints (6).

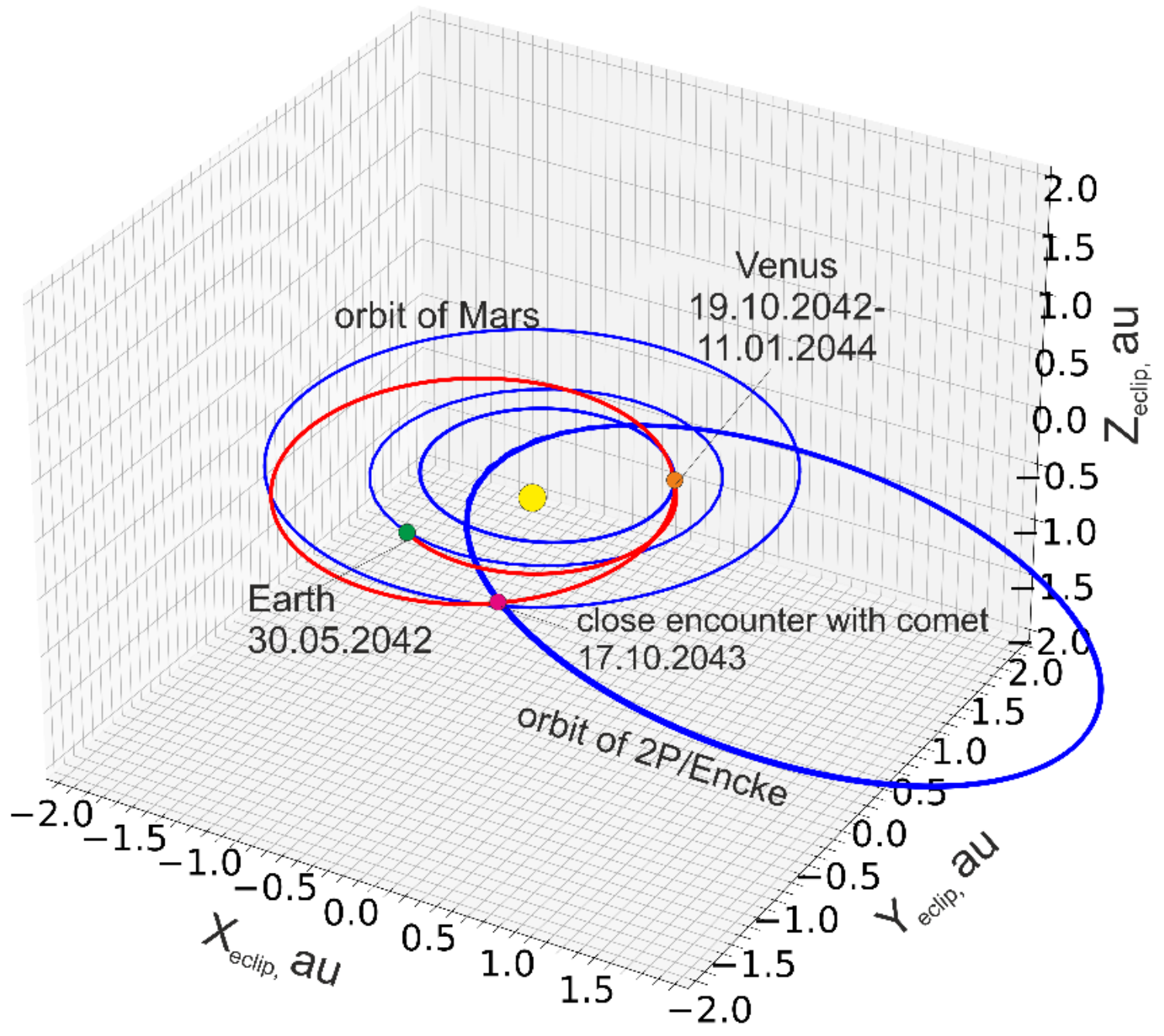

Fig. 7. Trajectory of the flight to Venus with the gravity assist maneuver and the passage of the 2P/Encke comet by the spacecraft in a resonant 1:2 orbit for mission launched at 30.05.2042

Certainly, if the mission scenario would be similar, for example, to the *Vega-1* and *Vega-2* missions, i.e. there would be no need to return the vehicle to Venus, it would not be difficult to find more suitable solutions for $\Delta V_0$, for example, the launch in 2039 allows to reach the 2P/Encke comet with the cost of 4.1 km/s for the Earth-Venus-2P/Encke scheme, and the flight by the Earth-Venus-2P/Encke scheme in 2034 allows to reach the comet with $\Delta V_0$ = 3.78 km/s. However, in the considered case the main goal was to determine such trajectories that would ensure the return of the spacecraft to Venus after its flyby, so within the framework of this problem our solutions are still the most favorable.

Thus, despite the fact that the cost value $\Delta V_0$ for the flight to Venus in the Earth-Venus-Encke-Venus scenario exceeds the established limit of 4.5 km/s, but due to the unique opportunity to achieve simultaneously several important goals, such as landing on Venus and the study of the comet, this flight scenario becomes significantly relevant.

## 6. Analysis of attainable landing points on the Venus surface within the considered scenarios

Take a closer look to the results in terms of ensuring landing of the lander in the specific area on the surface of Venus while along with asteroid flyby. For this purpose, let us analyze available places for landing on the surface of Venus, which can be achieved during a flight along the trajectories including impulse-free encounter with an asteroid.

In Table 3 the parameters of the landing circles are given for the launch windows in 2029 and 2031. These launch windows are chosen primarily because in 2029-2031 it is planned to launch the spacecraft as part of the *Venera-D* mission. As part of this project a scenario with a gravity assist maneuver may be carried out to ensure landing in the required area on the surface of the planet [12,26]. Notice that trajectories of the spacecraft with a flyby of asteroids presented in Table 3 are shown in Appendix B at Fig. B1-B.6.

Table 3. The parameters of the landing circles on the Venus surface resulting from the spacecraft flight on the trajectory, accompanied by impulse-free asteroid flyby (2029 - 2031).

| Launch year | Asteroid name | Landing date | $V_{rk}$, km/s | $\lambda_C$, deg | $\varphi_C$, deg | $\psi$, deg |
|---|---|---|---|---|---|---|

| | | | | | | $\theta$= -6, deg | $\theta$= -12, deg | $\theta$= -24, deg | $\theta$= -27, deg |
|---|---|---|---|---|---|---|---|---|---|
| 2029 | 2000 WB1 | 12.09.2030 | 6.16 | -143.6 | 16.1 | 64 | 73 | 90 | 95 |
| | 2006 KE89 | 27.09.2030 | 4.31 | -100.0 | 10.1 | 53 | 63 | 82 | 87 |
| 2031 | 2004 TG10 | 25.06.2032 | 6.19 | -133.3 | 61.4 | 64 | 73 | 91 | 95 |
| | 4341 Poseidon | 09.05.2032 | 3.56 | -82.4 | -19.0 | 47 | 57 | 78 | 83 |
| | 1996 FG3 | 04.07.2032 | 5.61 | -95.2 | 8.1 | 61 | 70 | 88 | 93 |
| | 2002 SY50 | 31.03.2032 | 10.00 | 165.4 | -10.3 | 78 | 86 | 101 | 105 |

*Note $V_{rk}$ - asymptotic velocity of the spacecraft on return to Venus; $\lambda_C$ - longitude of the landing circle center; $\varphi_C$ - latitude of the landing circle center; longitude and latitude of the landing circle center are given in the planetocentric coordinate system; $\psi$ - radius of the landing circle, $\theta$ - entry angle.*

The radius of the landing circle shown in the Table 3 can be calculated as follows [27]:

$$\psi = \arccos\left(\frac{\mu}{\mu + r_p \left(V_r\right)^2}\right) + \arcsin\frac{tg\theta_{Entry}\, p}{e\, r_{Entry}} + L / R_{Venus}\,, \qquad (7)$$

where $r_p ,p,e$ - radius of pericenter (km), semilatus rectum (km) and eccentricity of an approach hyperbolic trajectory of the spacecraft, $\theta_{Entry}, r_{Entry}$ - entry angle (deg) and planetocentric distance (km) from the planet center to the point of the spacecraft entry into the atmosphere; $L$ - shift of landing point relative to the point of entry into the atmosphere, km, $R_{Venus} = 6051\ km$ - mean radius of Venus, km. The value $L / R_{Venus}$ is determined by numerical integration of the equations of motion of the lander in the dense layers of the atmosphere of Venus. In this work the trajectories were integrated by Runge-Kutta method (8) 9 order, the model of Venusian atmosphere was adopted from [28].

Figure 8 shows a surface map of Venus on which the landing circles, which can be achieved by the spacecraft flight using the scheme Earth-Venus-asteroid-Venus, are drawn. The landing circles are plotted for an entry angle of –12 deg (Table 3).

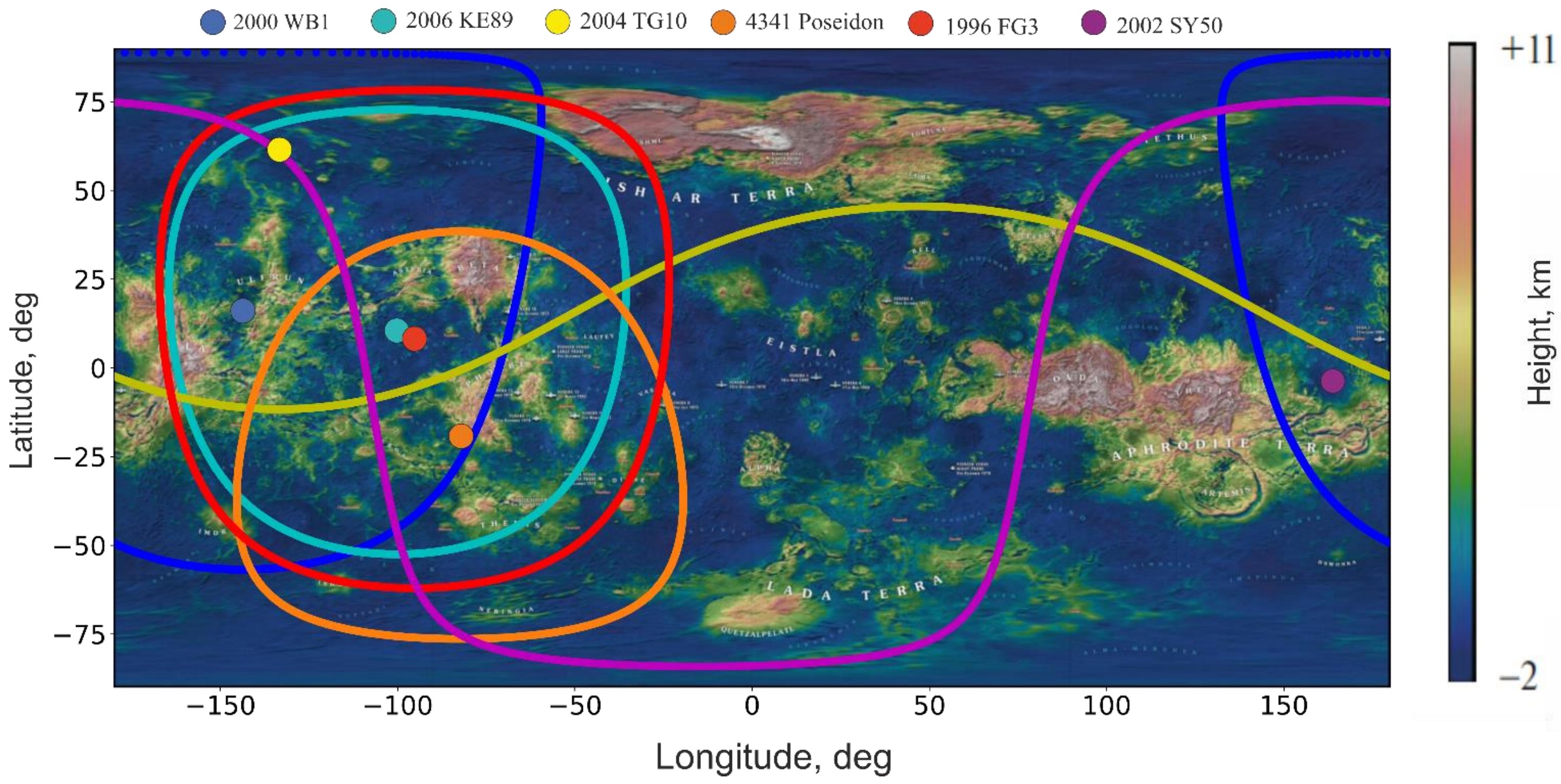

Fig. 8. Landing circles resulting from the spacecraft flight under the scheme Earth-Venus-asteroid-Venus; landing circles is considered for the launch of mission to Venus in 2029 and 2031.

It can be seen from Fig. 8 that passing the asteroids leads to the reduction of the number of available landing sites out of the whole set of attainable landing areas given in [12,29]. However, there may be a compromise between the choice of landing point, which has a sufficiently high scientific value, and the asteroid, the study of which also turns out to be possible.

In addition to Table 3, let us provide the results for the launch windows from 2032 to 2040 (see Table 4). This division into periods is based on the fact that the years from 2032 to 2040 are backups for the launch of the *Venera-D* mission [14,30] as well as other missions to Venus.

Table 4. The parameters of the landing circles on the Venus surface resulting from the spacecraft flight on the trajectory, accompanied by impulse-free asteroid flyby (2032 – 2040).

| Launch year | Asteroid name | Landing date | $V_{rk}$, km/s | $\lambda_C$, deg | $\varphi_C$, deg | $\psi$, deg | | | |
|---|---|---|---|---|---|---|---|---|---|
| | | | | | | $\theta$= -6, deg | $\theta$= -12, deg | $\theta$= -24, deg | $\theta$= -27, deg |
| 2032-2033 | 6063 Jason | 13.03.2034 | 3.97 | -156 | 1 | 50 | 60 | 80 | 85 |
| | 1685 Toro | 12.11.2033 | 8.29 | -173 | -14 | 73 | 81 | 97 | 101 |
| | 9162 Kwiila | 04.12.2033 | 4.63 | -110 | 0 | 55 | 65 | 84 | 88 |
| | 2P/Encke | 04.03.2034 | 6.70 | -20 | -38 | 66 | 75 | 92 | 97 |
| | 2003 UC20 | 19.03.2034 | 7.73 | 24 | -17 | 71 | 79 | 96 | 100 |
| | 139289 | 13.01.2034 | 8.84 | 160 | 78 | 75 | 83 | 98 | 102 |
| 2034-2035 | 1998 KN3 | 26.06.2035 | 4.62 | 19 | -32 | 55 | 65 | 84 | 88 |
| | 141484 | 12.07.2035 | 3.17 | 68 | -13 | 44 | 54 | 75 | 80 |
| | 308043 | 13.08.2035 | 3.85 | 155 | 75 | 49 | 60 | 79 | 84 |
| | 1865 Cerberus | 12.06.2035 | 8.86 | 166 | -6 | 75 | 83 | 98 | 102 |
| | 9162 Kwiila | 20.07.2035 | 6.85 | -78 | -38 | 67 | 76 | 93 | 97 |
| | 163243 | 18.08.2035 | 5.59 | 15 | 66 | 61 | 70 | 88 | 93 |
| | 4544 Xanthus | 12.10.2035 | 7.52 | -10 | 1 | 70 | 79 | 95 | 99 |
| 2036 | 159686 | 29.01.2037 | 7.70 | -156 | 27 | 71 | 79 | 96 | 100 |
| | 5731 Zeus | 05.05.2037 | 7.15 | -52 | -47 | 68 | 77 | 94 | 98 |
| | 154035 | 02.06.2037 | 8.19 | 33 | 20 | 72 | 81 | 97 | 101 |
| 2037-2038 | 139289 | 22.10.2038 | 5.33 | -11 | 36 | 59 | 69 | 87 | 92 |
| | 154276 | 18.09.2038 | 4.97 | -114 | -35 | 57 | 67 | 85 | 90 |
| | 141484 | 14.11.2038 | 4.91 | -128 | 41 | 57 | 66 | 85 | 90 |
| | 194268 | 14.12.2038 | 6.20 | -36 | -9 | 64 | 73 | 91 | 95 |
| | 4197 Morpheus | 28.12.2038 | 7.02 | 4 | 1 | 68 | 77 | 94 | 98 |
| | 184990 | 27.12.2038 | 6.99 | 4 | -6 | 68 | 76 | 93 | 98 |
| 2040 | 1686 Toro | 11.11.2041 | 7.93 | -167 | -19 | 71 | 80 | 96 | 100 |
| | 3554 Amun | 16.11.2041 | 6.81 | -152 | 65 | 67 | 76 | 93 | 97 |

Table 4 shows that the number of available asteroids increases for the launch dates from 2032 to 2040, compared to the previously studied launch dates in 2029 and 2031.

Let us present the result of an analysis of the total reachable landing regions for the launch dates from 2029 to 2040. To do this, let us map the coordinates of the centers of the circles given in Tables 3 and 4 and in Fig. 9.

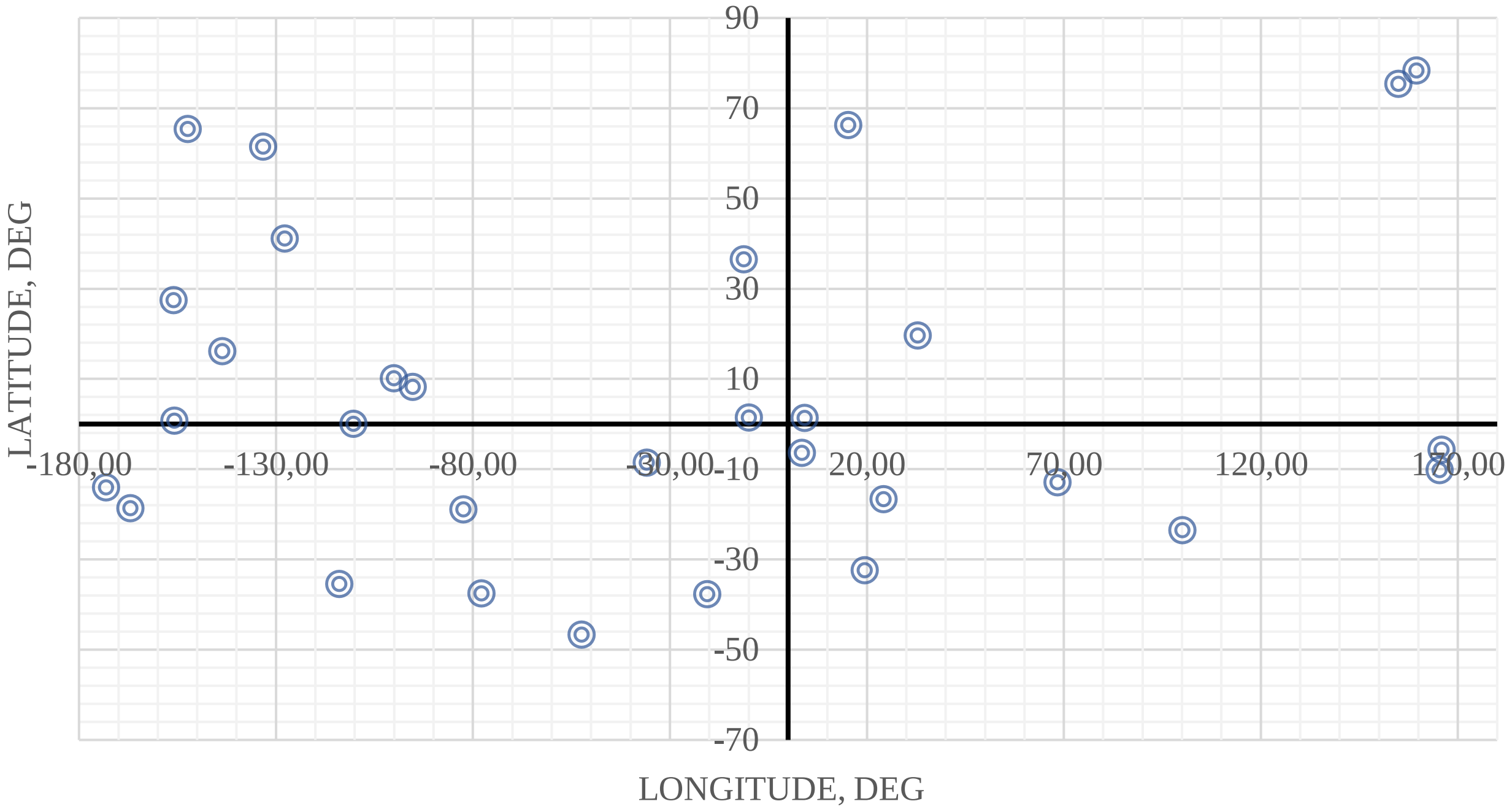

Fig. 9. Centers of landing circles in the planetocentric coordinate system (body fixed frame), obtained for the launch windows from 2029 to 2040.

Fig. 9 shows the coordinates of the centers of the circles of landing obtained in the implementation obtained in this study the trajectories of spacecraft flight to Venus with asteroid flyby in the launch windows from 2029 to 2040. Hence, we can see that most of the surface of Venus is achieved when implementing the scheme of the spacecraft flight with a flyby both Venus and the asteroid. Note that most of the centers of the landing circles are in the western longitudes in the northern hemisphere of Venus, however, since the landing circle has a non-zero radius the most of the surface of Venus even in this case is reachable. Let us show in Fig. 10 the total attainable area of landing on the surface of Venus, plotted considering the data of Tables 3 and 4 for the entry angle of -12 deg.

c

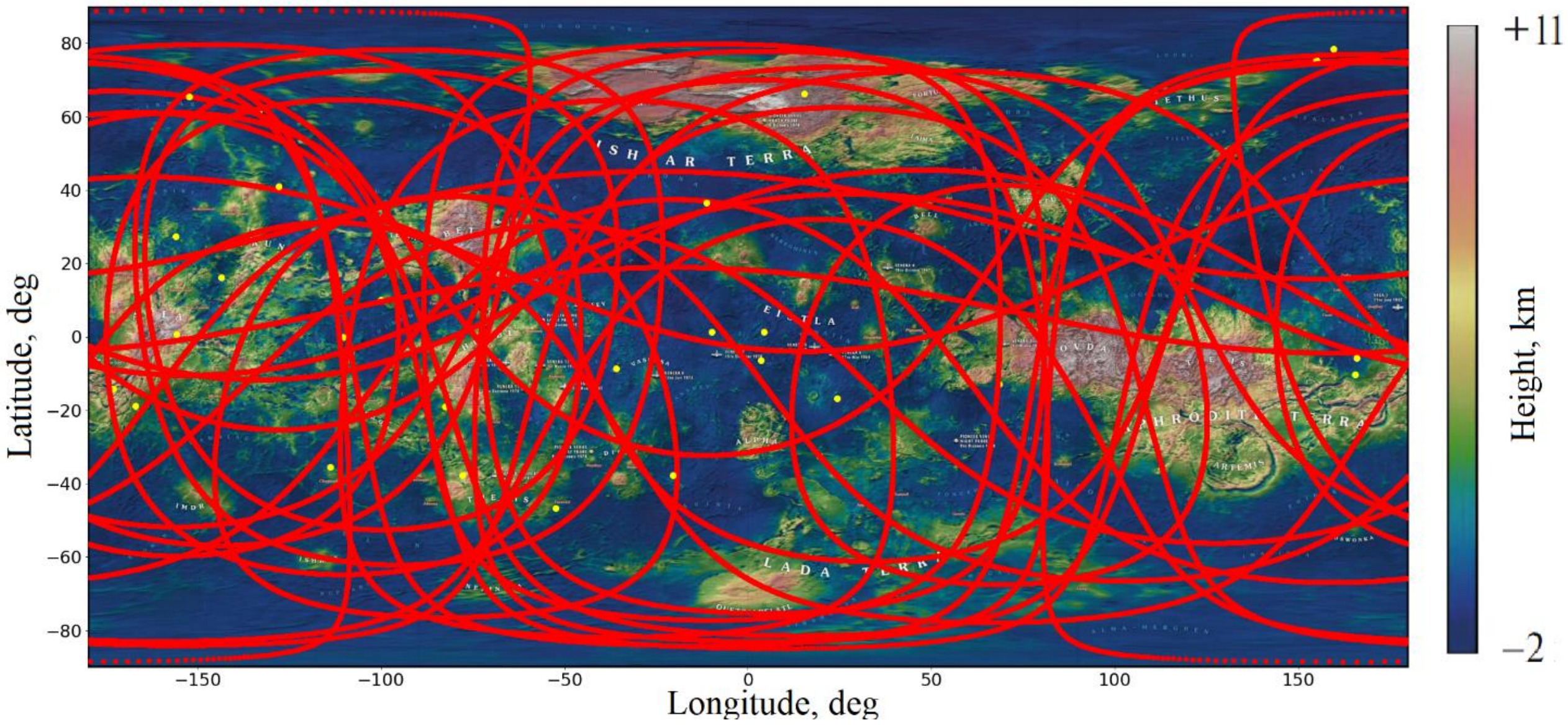

Fig. 10 The total attainable landing area (a set of red lines) when using schemes of the spacecraft flight, including the Venus and the asteroid flyby, and the subsequent return to Venus

Based on Fig. 10, it can be claimed that in almost any launch window for the flight to Venus there can be found such a trajectory of the spacecraft flight with the gravity assist maneuver, which includes an impulse-free encounter with an asteroid and at the same time satisfy the requirements to attain the selected landing site.

## Conclusions

In this paper, the problem of constructing a flight scheme to Venus with a flyby of an asteroid and a subsequent landing on its surface was considered. Let us highlight the most important points of the research results:

1. In the search for asteroids suitable for the study, 117 candidate asteroids from the NASA JPL catalog, whose diameter exceeds 1 km, were selected. However, when studying the possibility of their impulse-free passage by the spacecraft in the direct flight from the Earth to Venus for the launch dates from 2029 to 2050 it was impossible to find a trajectory of the flight, satisfying, firstly, the requirement of impulse-free flyby of an asteroid, and secondly, with a duration in the limits corresponding to the trajectories of the second semi-turn.

2. As a result of the search for optimal flight trajectories of the spacecraft, it was found that fulfillment of the restrictions on the impulse-free passage of both Venus and an asteroid was reached if the spacecraft after the gravity assist of Venus goes to a resonant 1:1 orbit. This dependence was established through numerous numerical simulations of the flight trajectories. Only in two cases the above conclusion was violated, in the case of comet 2P/Encke flyby and in the case of the flyby 6063 Jason asteroid. Though the latter may be explained by the fact that the encounter with 6063 Jason occurrs when the inclination of the transfer trajectory is the same as Venus' orbit to the ecliptic.
3. In the scenario of flight to Venus with a gravity assist and flight on a resonant 1:1 orbit it was found 35 asteroids, the passage of which satisfies all of the outlined constraints (6).
4. An example of a flight to Venus with a flyby of the 2P/Encke comet for a launch date in 2033 is shown, but it was turned out that such a mission is possible only if the launch impulse is at least 5.13 km/s.
5. Analysis of attainable landing areas in the implementation of trajectories of the spacecraft flight to Venus with an impulse-free encounter with an asteroid, has demonstrated that in each of the launch windows it can be found a trajectory that satisfying both the requirement for an impulse-free passage of the asteroid, and return the lander to the desired point on the surface of Venus after a revolution of the spacecraft along the resonant 1:1 orbit.

## APPENDIX A. PARAMETERS OF TRAJECTORIES TO VENUS WITH ASTEROID FLYBY

This appendix contains the parameters of the trajectories of the spacecraft flight to Venus with an impulse-free passage of the asteroid between two subsequent approaches of Venus for the launch dates from 2029 to 2050.

Table A.1 – Selected characteristics of the flight trajectories of the spacecraft to Venus with gravity assist maneuver and asteroid flyby

| Launch year | Designation of asteroid in JPL catalogueue | Celestial body name | Launch date | Date of Venus flyby | Date of asteroid flyby | Date of landing on the Venus surface | $\Delta V_0$, km/s | Pericenter height (at Venus flyby), $10^3$ km | Asymptotic velocity of spacecraft during Venus flyby, km/s | Asymptotic velocity of spacecraft during asteroid flyby, km/s | Inclination of spacecraft orbit on Venus-Venus route to ecliptic, deg |
|---|---|---|---|---|---|---|---|---|---|---|---|
| 2029 | 2153195 | 153195 (2000 WB1) | 19.10.2029 | 30.01.2030 | 02.06.2030 | 12.09.2030 | 3.83 | 8.67 | 6.16 | 28.72 | 5.49 |
| | 2184990 | 184990 (2006 KE89) | 28.10.2029 | 14.02.2030 | 08.04.2030 | 27.09.2030 | 3.87 | 12.23 | 4.31 | 42.52 | 3.67 |
| 2031 | 3256324 | 2004 TH10 | 19.05.2031 | 19.08.2031 | 06.10.2031 | 31.03.2032 | 4.21 | 10.83 | 10.00 | 34.04 | 5.71 |
| | 2004341 | 4341 Poseidon (1987 KF) | 19.05.2031 | 13.11.2031 | 09.01.2032 | 25.06.2032 | 4.31 | 0.50 | 6.19 | 19.38 | 11.77 |
| | 2154276 | 154276 (2002 SY50) | 21.05.2031 | 27.09.2031 | 25.12.2031 | 09.05.2032 | 3.82 | 12.67 | 3.56 | 17.73 | 2.39 |
| | 2175706 | 1996 FG3 | 23.05.2031 | 22.11.2031 | 19.03.2032 | 04.07.2032 | 3.76 | 24.24 | 5.61 | 6.66 | 4.11 |
| 2032-33 | 2006063 | 6063 Jason (1984 KB) | 21.11.2032 | 09.05.2033 | 05.11.2033 | 14.03.2034 | 3.96 | 0.50 | 3.97 | 17.49 | 3.40 |
| | 2001685 | 1685 Toro (1948 OA) | 23.12.2032 | 01.04.2033 | 31.07.2033 | 12.11.2033 | 4.09 | 16.74 | 8.29 | 11.70 | 5.55 |
| | 2009162 | 9162 Kwiila (1987 OA) | 29.12.2032 | 23.04.2033 | 07.07.2033 | 04.12.2033 | 3.66 | 28.82 | 4.63 | 17.17 | 3.40 |
| | 1000025 | 2P/Encke | 29.01.2033 | 02.06.2033 | 22.08.2033 | 13.01.2034 | 5.13 | 14.08 | 8.88 | 30.72 | 12.20 |
| | 2363505 | 2003 UC20 | 31.01.2033 | 23.07.2033 | 20.10.2033 | 04.03.2034 | 3.86 | 18.70 | 6.70 | 8.08 | 5.58 |
| | 2139289 | 139289 (2001 KR1) | 15.02.2033 | 06.08.2033 | 21.01.2034 | 19.03.2034 | 3.88 | 33.51 | 7.73 | 38.96 | 2.09 |
| 2034 | 2387505 | 1998 KN3 | 28.04.2034 | 14.11.2034 | 15.01.2035 | 26.06.2035 | 4.18 | 0.50 | 4.62 | 35.04 | 1.00 |
| | 2141484 | 141484 (2002 DB4) | 04.06.2034 | 29.11.2034 | 11.03.2035 | 12.07.2035 | 3.86 | 1.35 | 3.17 | 15.09 | 2.16 |
| | 3256325 | 2005 TG10 | 03.07.2034 | 31.12.2034 | 23.03.2035 | 13.08.2035 | 3.88 | 15.87 | 3.85 | 34.05 | 8.02 |
| | 2001865 | 1865 Cerberus (1971 UA) | 17.07.2034 | 31.10.2034 | 14.01.2035 | 12.06.2035 | 4.05 | 26.01 | 8.82 | 13.66 | 2.70 |
| | 2009162 | 9162 Kwiila (1987 OA) | 21.07.2034 | 08.12.2034 | 18.04.2035 | 21.07.2035 | 3.99 | 4.00 | 6.85 | 22.21 | 3.91 |
| | 2163243 | 163243 (2002 FB3) | 31.07.2034 | 05.01.2035 | 21.02.2035 | 18.08.2035 | 4.26 | 6.22 | 5.58 | 25.94 | 10.01 |
| | 2004544 | 4544 Xanthus (1989 FB) | 15.09.2034 | 02.03.2035 | 23.06.2035 | 13.10.2035 | 3.80 | 17.15 | 7.53 | 12.78 | 3.10 |
| 2036 | 2159686 | 159686 (2002 LB6) | 06.03.2036 | 18.06.2036 | 23.08.2036 | 29.01.2037 | 3.84 | 26.20 | 7.71 | 20.83 | 2.78 |
| | 2005731 | 5731 Zeus (1988 VP4) | 06.03.2036 | 22.09.2036 | 27.10.2036 | 04.05.2037 | 4.00 | 9.21 | 7.15 | 14.10 | 5.21 |
| | 2154035 | 154035 (2002 CV59) | 30.04.2036 | 20.10.2036 | 12.04.2037 | 01.06.2037 | 3.90 | 7.91 | 8.19 | 38.73 | 7.05 |
| 2037-38 | 2139290 | 139289 (2001 KR1) | 26.09.2037 | 11.03.2038 | 15.04.2038 | 22.10.2038 | 3.87 | 0.97 | 5.33 | 37.05 | 3.80 |
| | 2154277 | 154276 (2002 SY50) | 21.10.2037 | 06.02.2038 | 05.07.2038 | 18.09.2038 | 3.82 | 30.89 | 4.97 | 23.00 | 5.10 |
| | 2194268 | 2001 UY4 | 18.11.2037 | 04.05.2038 | 13.07.2038 | 14.12.2038 | 3.80 | 24.66 | 6.21 | 29.77 | 4.36 |
| | 2004197 | 4197 Morpheus (1982 TA) | 28.11.2037 | 17.05.2038 | 21.10.2038 | 28.12.2038 | 3.92 | 34.94 | 7.02 | 23.52 | 3.09 |
| | 2184991 | 184990 (2006 KE89) | 29.11.2037 | 17.05.2038 | 01.12.2038 | 27.12.2038 | 3.91 | 30.92 | 6.99 | 47.44 | 3.86 |
| 2040 | 2001686 | 1686 Toro (1948 OA) | 20.12.2040 | 31.03.2041 | 29.07.2041 | 11.11.2041 | 4.06 | 12.29 | 7.95 | 12.01 | 6.70 |
| | 2003554 | 3554 Amun (1986 EB) | 20.12.2040 | 05.04.2041 | 30.06.2041 | 16.11.2041 | 4.33 | 0.50 | 6.80 | 21.30 | 8.45 |
| 2042 | 2154278 | 154276 (2002 SY50) | 08.06.2042 | 06.12.2042 | 18.02.2043 | 19.07.2043 | 3.86 | 8.42 | 2.96 | 22.55 | 1.81 |
| | 2141485 | 141484 (2002 DB4) | 15.06.2042 | 15.12.2042 | 28.02.2043 | 28.07.2043 | 3.87 | 6.65 | 3.12 | 19.72 | 7.27 |
| | 2138013 | 138013 (2000 CN101) | 30.06.2042 | 16.12.2042 | 13.05.2043 | 29.07.2043 | 4.00 | 0.50 | 3.65 | 20.60 | 3.30 |
| | 2184992 | 184990 (2006 KE89) | 30.06.2042 | 19.01.2043 | 20.03.2043 | 31.08.2043 | 4.09 | 5.21 | 5.78 | 36.33 | 8.31 |
| | 2003554 | 3555 Amun (1986 EB) | 25.07.2042 | 16.01.2043 | 27.05.2043 | 29.08.2043 | 3.83 | 28.34 | 5.02 | 18.42 | 5.05 |
| | 2001862 | 1862 Apollo (1932 HA) | 06.08.2042 | 23.10.2042 | 31.01.2043 | 05.06.2043 | 4.40 | 2.66 | 11.49 | 25.30 | 8.34 |
| | 2184993 | 184990 (2006 KE89) | 15.08.2042 | 19.01.2043 | 20.03.2043 | 01.09.2043 | 3.99 | 21.53 | 5.81 | 36.36 | 8.30 |
| | 2066146 | 1998 TU3 | 23.08.2042 | 22.01.2043 | 06.05.2043 | 04.09.2043 | 4.12 | 4.63 | 6.38 | 21.04 | 3.34 |
| | 2055532 | 55532 (2001 WG2) | 25.08.2042 | 23.12.2042 | 09.03.2043 | 04.08.2043 | 4.33 | 6.09 | 6.39 | 32.85 | 7.59 |
| | 2004769 | 4769 Castalia (1989 PB) | 30.08.2042 | 06.03.2043 | 01.06.2043 | 17.10.2043 | 4.08 | 13.07 | 8.49 | 13.72 | 4.20 |
| | 1000026 | 2P/Encke | 30.09.2042 | 11.04.2043 | 17.07.2043 | 22.11.2043 | 5.25 | 0.51 | 11.85 | 25.10 | 11.10 |

| | | | | | | | | | | | | |
|---|---|---|---|---|---|---|---|---|---|---|---|---|
| | 2006064 | 6064 Jason (1984 KB) | 20.10.2042 | 03.04.2043 | 20.09.2043 | 14.11.2043 | 4.11 | 30.64 | 9.40 | 15.29 | 2.54 |
| 2044 | 2363506 | 2004 UC20 | 24.03.2044 | 13.09.2044 | 05.11.2044 | 26.04.2045 | 3.66 | 24.69 | 6.01 | 11.74 | 3.10 |
| | 2003554 | 3556 Amun (1986 EB) | 29.06.2044 | 07.12.2044 | 25.04.2045 | 20.07.2045 | 4.49 | 5.59 | 11.81 | 16.09 | 7.63 |
| 2045 | 2004486 | Mithra (1987 SB) | 04.10.2045 | 14.03.2046 | 12.05.2046 | 25.10.2046 | 3.81 | 2.52 | 5.47 | 15.14 | 3.04 |
| | 2066147 | 1999 TU3 | 20.10.2045 | 30.03.2046 | 17.08.2046 | 09.11.2046 | 3.61 | 30.01 | 4.97 | 15.66 | 2.56 |
| 2047 | 2153196 | 153195 (2000 WB1) | 26.05.2047 | 13.11.2047 | 13.01.2048 | 24.06.2048 | 3.67 | 18.25 | 4.99 | 29.85 | 5.90 |
| | 2003753 | 3753 Cruithne (1986 TO) | 07.06.2047 | 23.08.2047 | 23.10.2047 | 04.04.2048 | 4.30 | 0.50 | 8.89 | 20.05 | 11.17 |
| 2048-49 | 2139345 | 139345 (2001 KA67) | 08.11.2048 | 22.04.2049 | 23.09.2049 | 03.12.2049 | 4.42 | 0.50 | 7.46 | 23.78 | 6.48 |
| | 2154279 | 154276 (2002 SY50) | 12.12.2048 | 13.06.2049 | 22.08.2049 | 24.01.2050 | 3.87 | 15.37 | 4.45 | 20.96 | 4.13 |
| | 2001687 | 1687 Toro (1948 OA) | 15.12.2048 | 29.03.2049 | 26.07.2049 | 08.11.2049 | 4.07 | 6.14 | 7.80 | 12.71 | 8.83 |
| | 2001687 | 1688 Toro (1948 OA) | 01.01.2049 | 28.03.2049 | 22.07.2049 | 08.11.2049 | 4.26 | 0.50 | 8.54 | 13.72 | 11.22 |
| | 2154280 | 154276 (2002 SY50) | 07.01.2049 | 15.06.2049 | 21.08.2049 | 26.01.2050 | 3.94 | 5.22 | 4.57 | 21.15 | 4.25 |
| 2050 | 2138014 | 138013 (2000 CN101) | 22.08.2050 | 08.02.2051 | 06.06.2051 | 20.09.2051 | 3.77 | 18.42 | 6.55 | 23.80 | 5.47 |
| | 2003360 | 3360 Syrinx (1981 VA) | 11.06.2050 | 10.12.2050 | 10.03.2051 | 22.07.2051 | 3.85 | 13.44 | 3.01 | 25.70 | 7.39 |
| | 2003360 | 3360 Syrinx (1981 VA) | 10.08.2050 | 28.01.2051 | 04.03.2051 | 10.09.2051 | 3.79 | 8.81 | 5.91 | 28.70 | 4.75 |

# APPENDIX B. TRAJECTORIES OF THE SPACECRAFT FLIGHT TO VENUS WITH AN ASTEROID ENCOUNTER AFTER VENUS FLYBY

This appendix contains the picture of trajectories of the spacecraft to the asteroids outlined in Table 3.

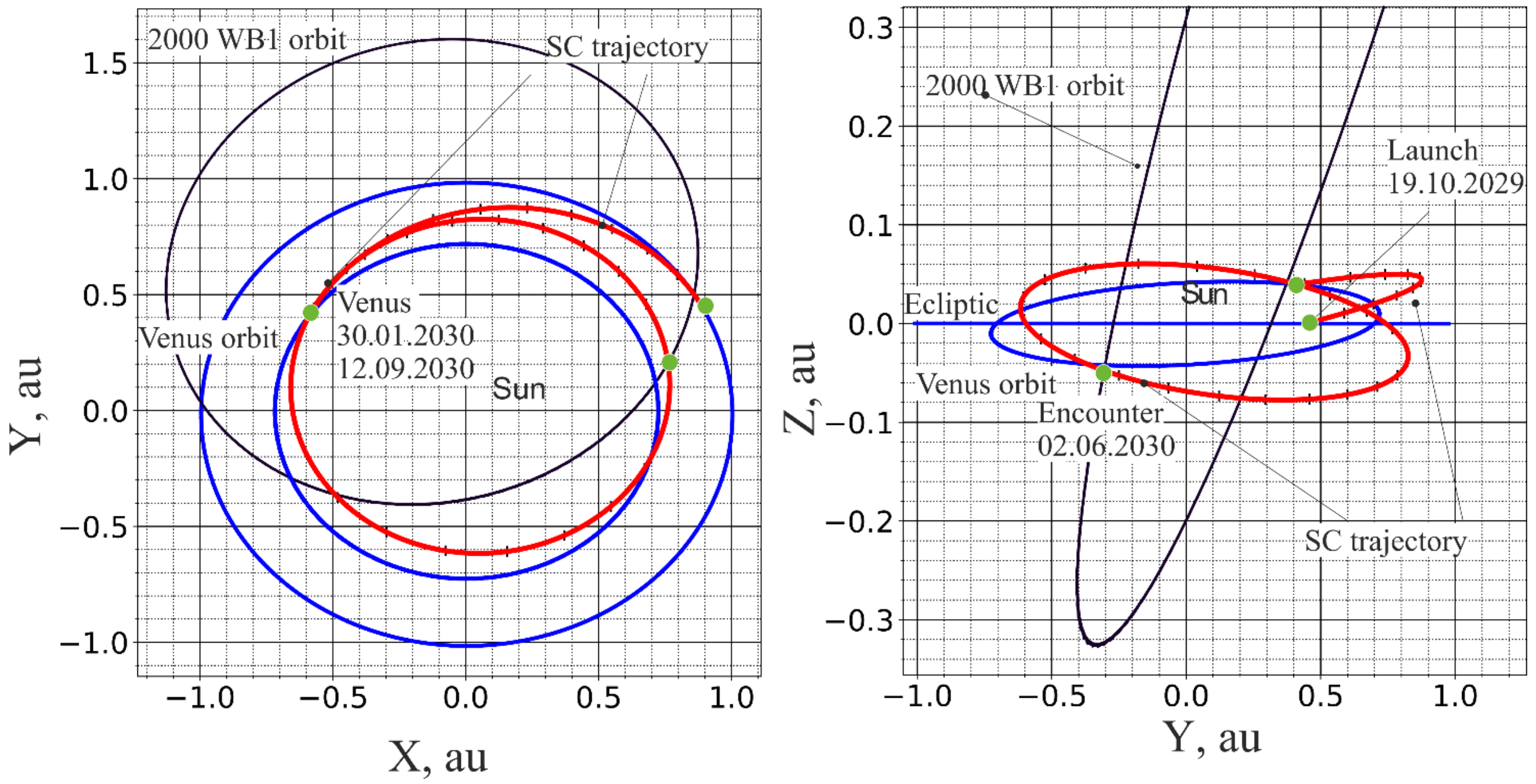

Fig. B.1 Trajectory of spacecraft (SC) to Venus with flyby of asteroid 2000 WB1 after Venus gravity assist and subsequent landing at 12.09.2030. X, Y, Z are the axis of heliocentric ecliptic frame

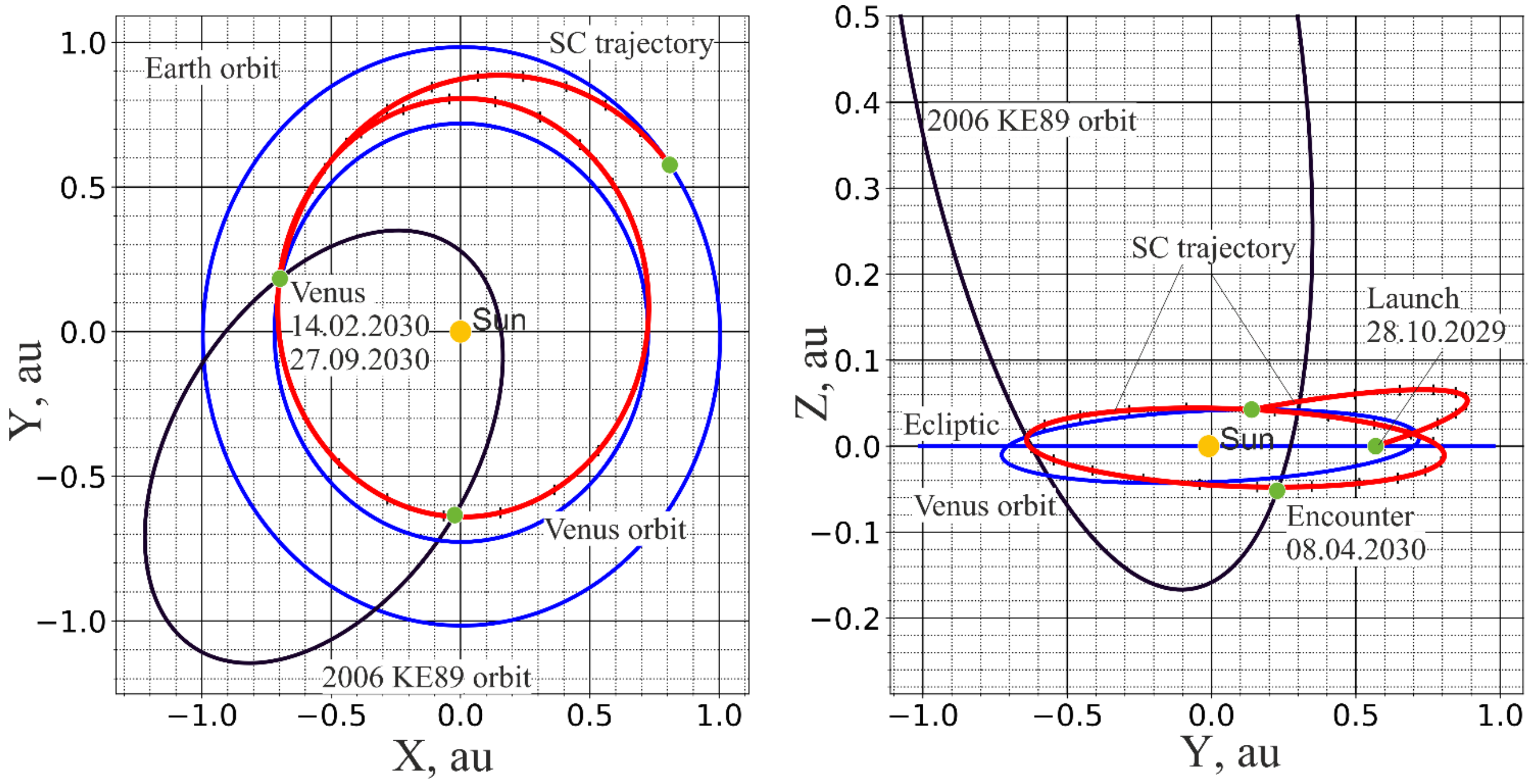

Fig. B.2 Trajectory of spacecraft to Venus with flyby of asteroid 2000 WB1 after Venus gravity assist and subsequent landing at 27.09.2030

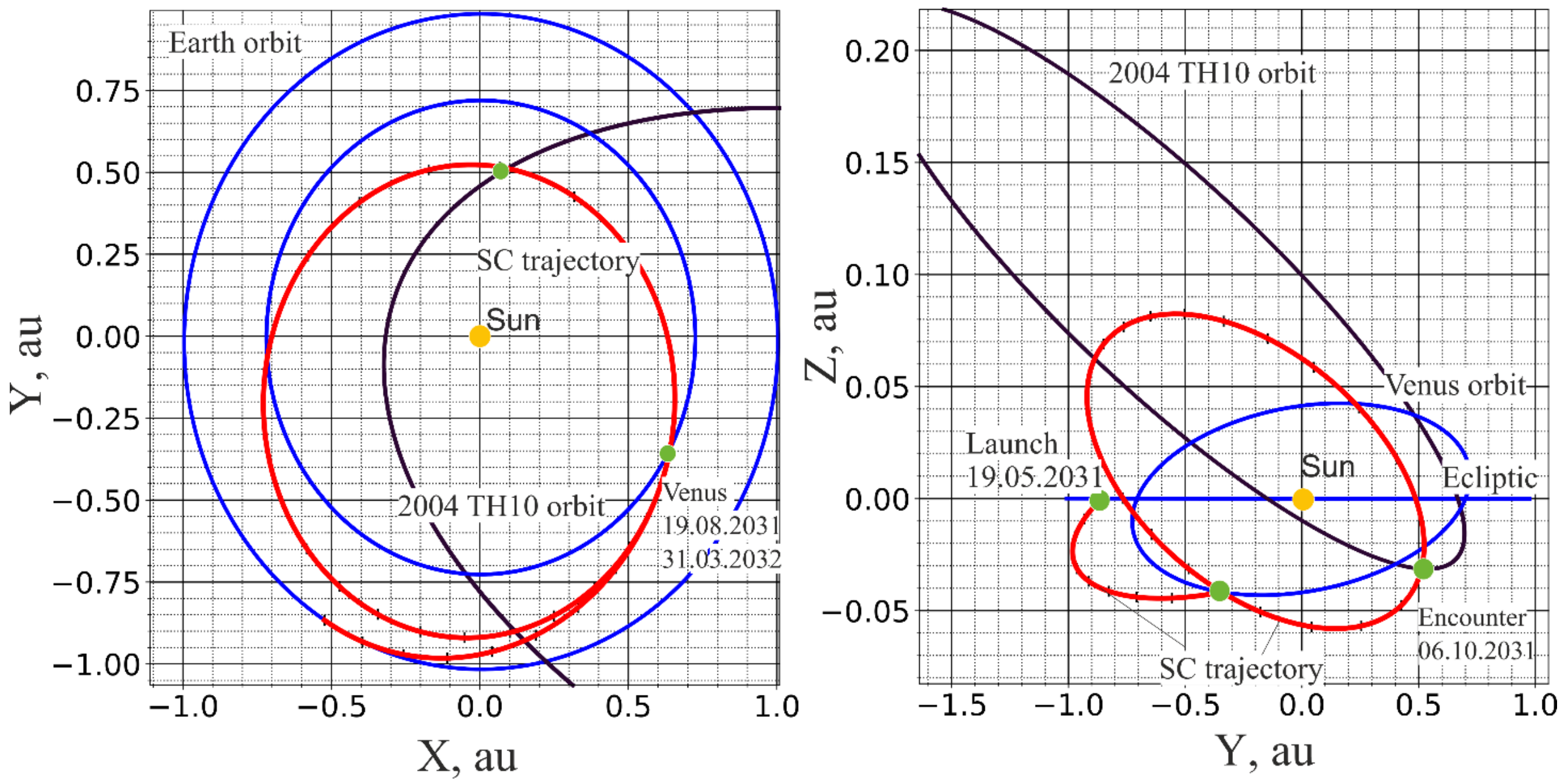

Fig. B.3 Trajectory of spacecraft to Venus with flyby of asteroid 2004 TH10 after Venus gravity assist and subsequent landing at 31.03.2032

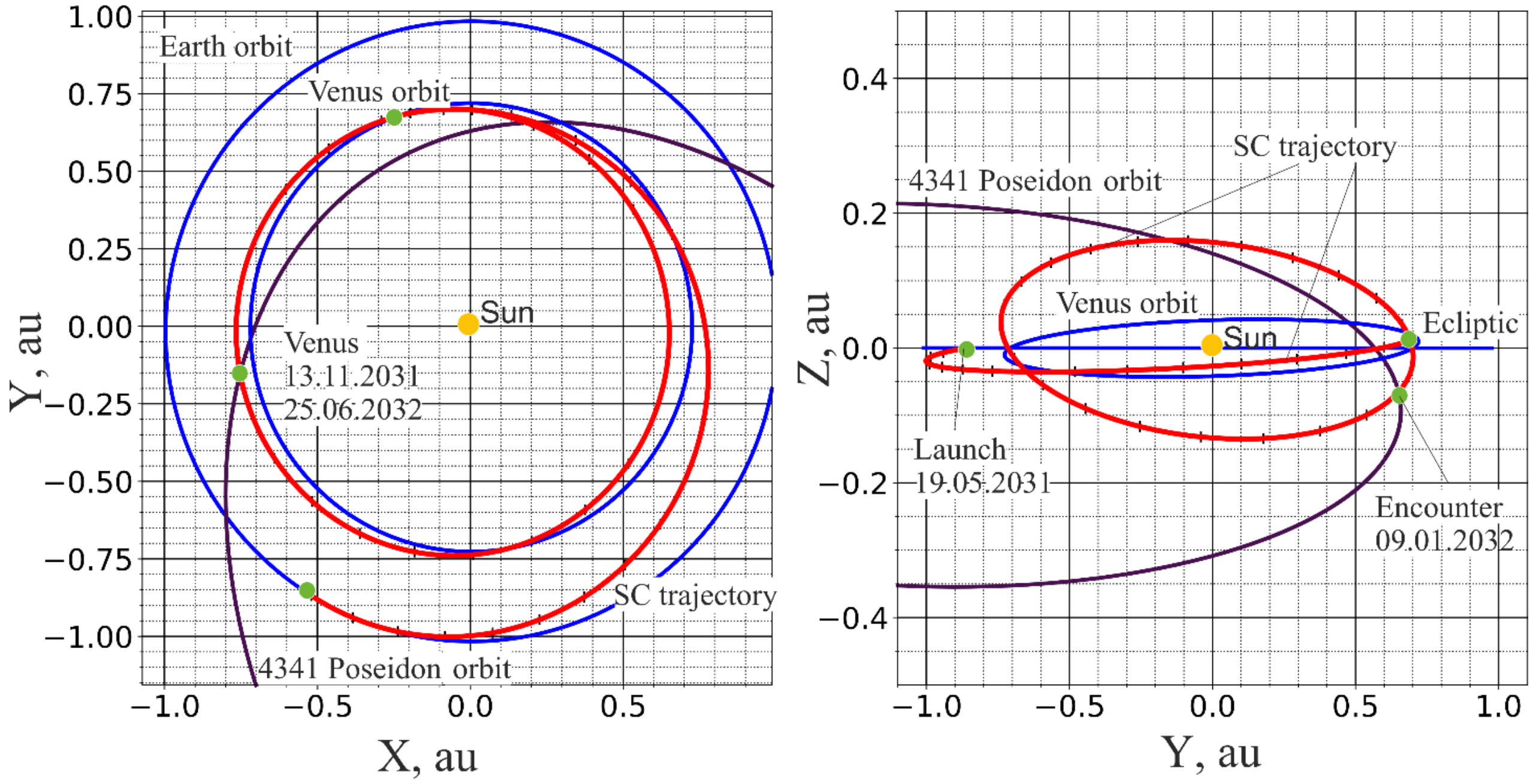

Fig. B.4 Trajectory of spacecraft to Venus with flyby of asteroid 4341 Poseidon after Venus gravity assist and subsequent landing at 25.06.2032

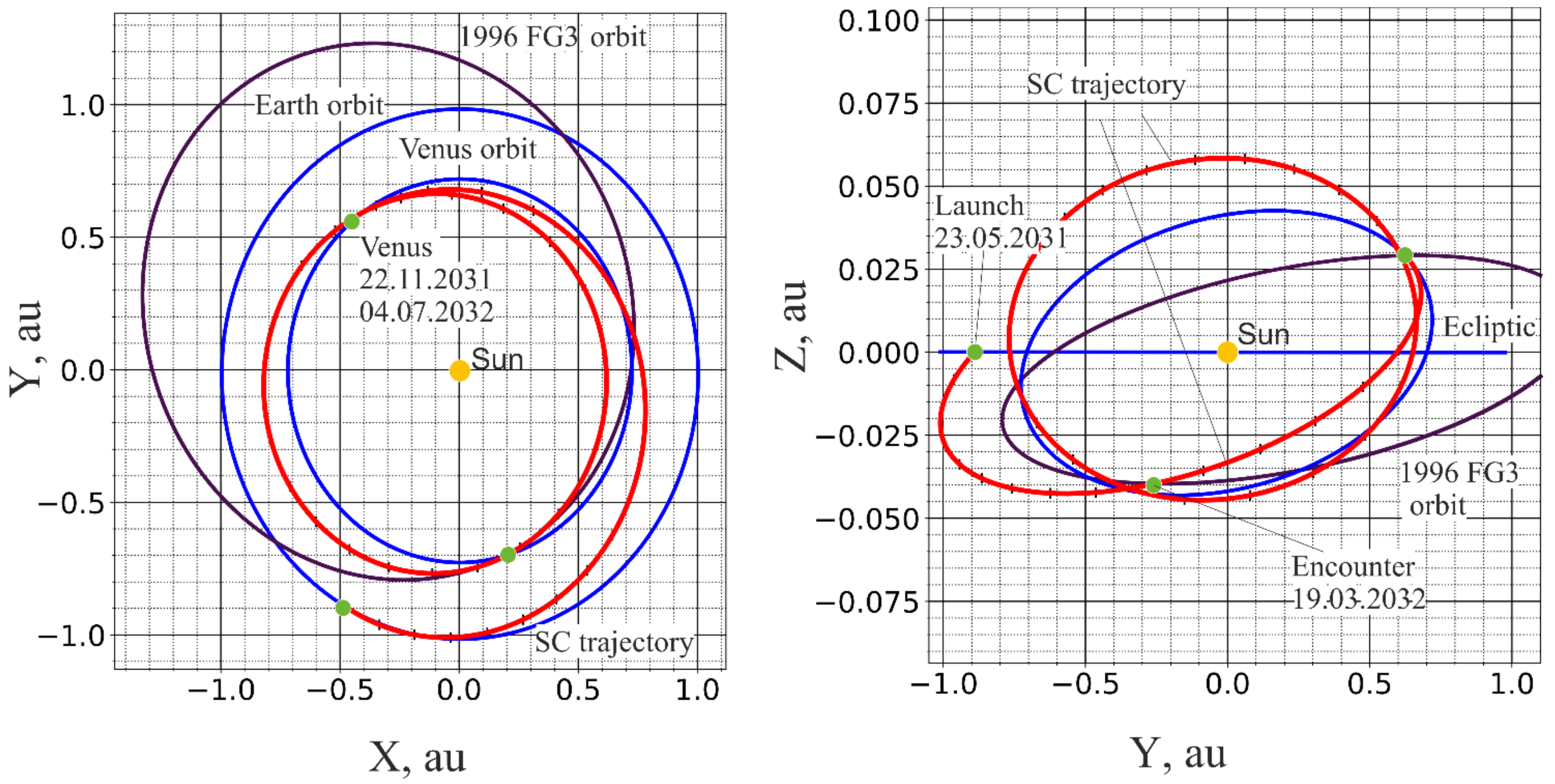

Fig. B.5 Trajectory of spacecraft to Venus with flyby of asteroid 1996 FG3 after Venus gravity assist and subsequent landing at 04.07.2032

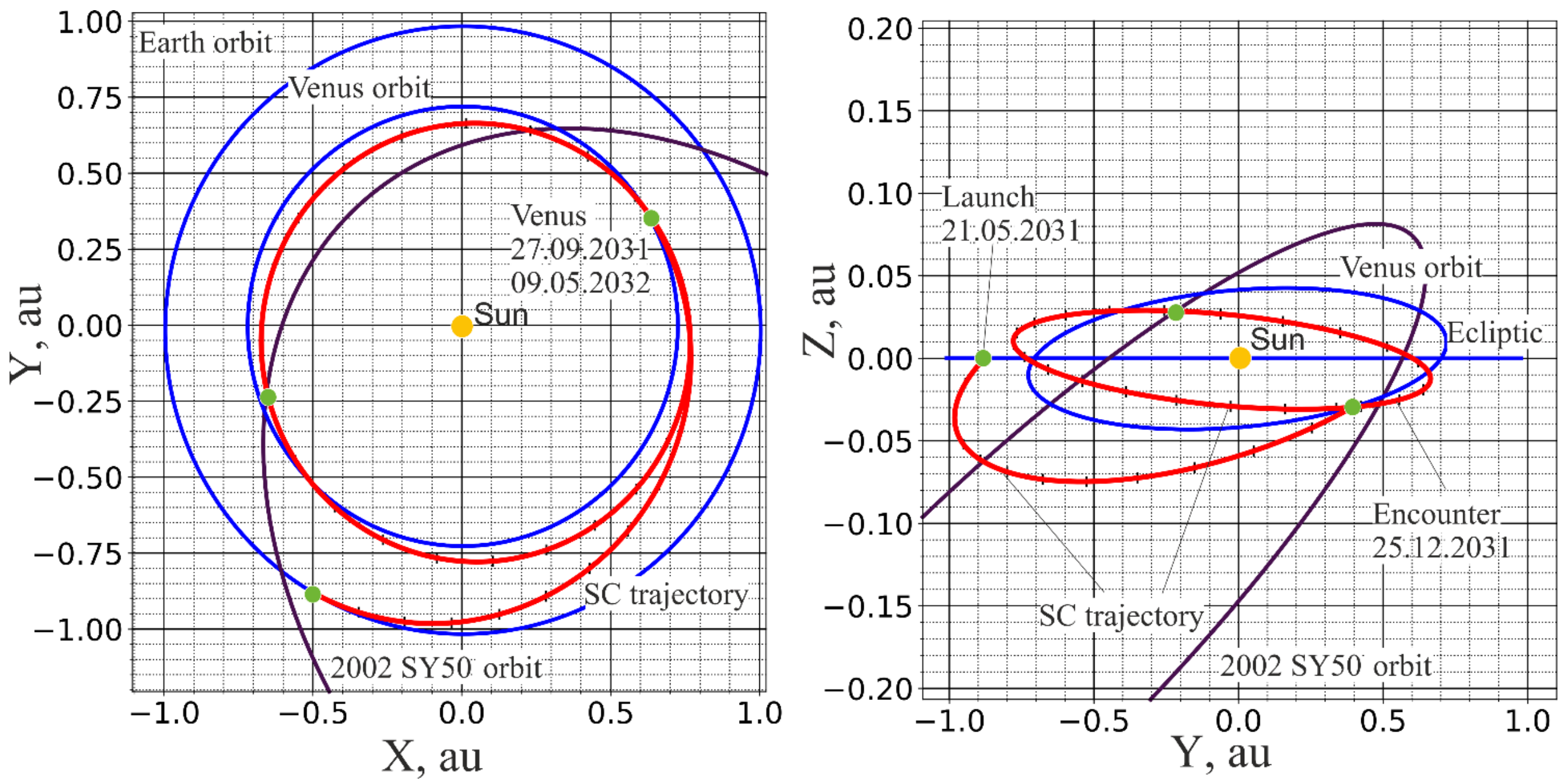

Fig. B.6 Trajectory of spacecraft to Venus with flyby of asteroid 2002 SY50 after Venus gravity assist and subsequent landing at 09.05.2032